\documentclass[aps,prb,10pt,twocolumn,superscriptaddress,floatfix]{revtex4-1}
\usepackage{graphicx,graphics}
\usepackage{dcolumn}
\usepackage{amsmath,amssymb,amsfonts}
\usepackage{latexsym,verbatim}
\usepackage{bm}
\usepackage{color}
\usepackage{textcomp}
\usepackage{gensymb}
\usepackage{ulem}
\usepackage[percent]{overpic}
\usepackage[breaklinks=true,colorlinks,citecolor=blue,linkcolor=blue,urlcolor=blue]{hyperref}
\usepackage{array}
\DeclareMathOperator\re{Re}
\DeclareMathOperator\im{Im}

\DeclareMathOperator\tr{Tr}
\begin{document}
\title{Optical and plasmonic properties of twisted bilayer graphene: Impact of inter-layer tunneling asymmetry and ground-state charge inhomogeneity}
\author{Pietro Novelli}
\email{pietro.novelli@sns.it}
\affiliation{NEST, Scuola Normale Superiore, Piazza dei Cavalieri 7, I-56126 Pisa, Italy}
\affiliation{Istituto Italiano di Tecnologia, Graphene Labs, Via Morego 30, I-16163 Genova, Italy}
\author{Iacopo Torre}
\affiliation{ICFO-Institut de Ci\`{e}ncies Fot\`{o}niques, The Barcelona Institute of Science and Technology, Av. Carl Friedrich Gauss 3, 08860 Castelldefels (Barcelona),~Spain}
\author{Frank H.L. Koppens}
\affiliation{ICFO-Institut de Ci\`{e}ncies Fot\`{o}niques, The Barcelona Institute of Science and Technology, Av. Carl Friedrich Gauss 3, 08860 Castelldefels (Barcelona),~Spain}
\affiliation{ICREA-Instituci\'{o} Catalana de Recerca i Estudis Avançats, Passeig Llu\'{i}s Companys 23, 08010 Barcelona, Spain}
\author{Fabio Taddei}
\affiliation{NEST, Istituto Nanoscienze-CNR, and Scuola Normale Superiore, Piazza dei Cavalieri 7, I-56126 Pisa, Italy}
\author{Marco Polini}
\affiliation{Dipartimento di Fisica dell'Universit\`a di Pisa, Largo Bruno Pontecorvo 3, I-56127 Pisa, Italy}
\affiliation{School of Physics \& Astronomy, University of Manchester, Oxford Road, Manchester M13 9PL, United Kingdom}
\affiliation{Istituto Italiano di Tecnologia, Graphene Labs, Via Morego 30, I-16163 Genova, Italy}
\begin{abstract}
Twisted bilayer graphene (TBG) at twist angles $\theta \approx 1\degree$ has recently attracted a great deal of interest for its rich transport phenomenology. 
We present a theoretical study of the local optical conductivity, plasmon spectra, and thermoelectric properties of TBG at different filling factors and twist angles $\theta$.
Our calculations are based on the electronic band structures obtained from a continuum model that has two tunable parameters, $u_0$ and $u_1$, which parametrize the intra-sublattice inter-layer and inter-sublattice inter-layer tunneling rate, respectively. In this Article we focus on two key aspects:  
i) we study the dependence of our results on the value of $u_0$, exploring the whole range $0\leq u_0\leq u_1$;
ii) we take into account effects arising from the intrinsic charge density inhomogeneity  present in TBG, by calculating the band structures within the self-consistent Hartree approximation.
At zero filling factor, i.e.~at the charge neutrality point, the optical conductivity is quite sensitive to the value of $u_0$ and twist angle, whereas the charge inhomogeneity brings about only modest corrections. 
On the other hand, away from zero filling, static screening dominates and the optical conductivity is appreciably affected by the charge inhomogeneity, the largest effects being seen on the intra-band contribution to it. 
These findings are also reflected by the plasmonic spectra. 
We compare our results with existing ones in the literature, where effects i) and ii) above have not been studied systematically. 
As natural byproducts of our calculations, we obtain the Drude weight and Seebeck coefficient. 
The former displays an enhanced particle-hole asymmetry stemming from the inhomogeneous ground-state charge distribution. 
The latter is shown to display a broad sign-changing feature even at low temperatures ($\approx 5~{\rm K}$) due to the reduced slope of the bands, as compared to those of single-layer graphene. 
\end{abstract}
\maketitle
\section{Introduction}
\label{sect:intro}
Twisted bilayer graphene (TBG)\cite{lopes_prl_2007,shallcross_prl_2008,mele_prb_2010,shallcross_prb_2010,li_naturephys_2010,bistritzer_prb_2010,bistritzer_pnas_2011,lopes_prb_2012} is a system consisting of two graphene sheets stacked one on top of each other, with a relative rotation of the crystal axes quantified by the twist angle $\theta$. 

The physics of the two-dimensional (2D) electron system roaming in TBG with twist angles $\theta \lesssim 3\degree$ is dominated by a triangular moir\'e pattern of periodicity $\approx a/\theta [{\rm rad}]$, where $a \approx 0.246~{\rm nm}$ is the lattice constant of monolayer graphene. 
In this case, the energy spectrum is well described by a continuum model~\cite{bistritzer_pnas_2011} which accounts for the long-range moir\'e modulations of the inter-layer tunneling amplitudes. 
At twist angles of  $\approx 1\degree$, the continuum model~\cite{bistritzer_pnas_2011} predicts the existence of a pair of remarkably flat bands, with a bandwidth of less than $20~{\rm meV}$, close to the charge neutrality point (CNP), where the Fermi level of an undoped sample lies. 
These bands exhibit a linear dispersion around the moir\'e Brillouin zone (MBZ) corners, with a twist-angle-dependent Fermi velocity $\hbar v_{\rm F}(\theta)$. 
The largest angle $\theta^{\star}$ satisfying $\hbar v_{{\rm F}}(\theta^{\star}) = 0$ is called the first {\it magic angle} (or simply {\it magic angle})~\cite{bistritzer_pnas_2011}. 

For systems with twist angles close to the magic one, the bandwidth of the flat bands reaches its minimum, ensuring a large density of states and strengthened electron-electron (e-e) interactions. In this regime ($\theta \sim 1.05\degree$), a plethora of intriguing phenomena have been recently observed, 
including superconductivity\cite{cao_nature_2018,yankowitz_science_2019,lu_nature_2019,stepanov_arxiv_2019}, 
correlated insulating states\cite{cao_nature_2018a,lu_nature_2019,yankowitz_science_2019,kerelsky_nature_2019,choi_natphys_2019,stepanov_arxiv_2019},
ferromagnetism\cite{sharpe_science_2019,lu_nature_2019},
charge-ordered states\cite{jiang_nature_2019,xie_nature_2019},
and a quantized anomalous Hall effect\cite{serlin_arxiv_2019, tomarken_prl_2019}.

The unique features of the low-energy spectrum of TBG also manifest in its optical properties, as showed experimentally e.g.~in Refs.~\onlinecite{hesp_arxiv_2019,utama_arxiv_2019}. 
In Ref.~\onlinecite{hesp_arxiv_2019}, in particular, the dispersion relation of collective excitations in TBG with $\theta \approx 1.35\degree$ has been directly measured through scanning near-field optical microscopy. 
This work unveiled the crucial role of the inter-layer tunneling amplitude in the determination of the optical and plasmonic properties of TBG. 
More precisely, the electronic band structures calculated using continuum models are influenced by two important parameters: i) the inter-layer tunnelling rate $u_0$ in the AA regions (which will be therefore dubbed ``intra-sublattice" tunnelling) and ii) the inter-layer tunnelling rate $u_1$ in the AB/BA regions (which will be therefore dubbed ``inter-sublattice" tunnelling). 
Comparing theoretical results with experimental data~\cite{hesp_arxiv_2019}, it was suggested that $u_0$ can be much smaller than $u_1$ in significant areas of real samples. 
For the sake of clarity, we remind the reader that in the literature the cases $u_0=u_1$ or $u_0\lesssim u_1$ are often studied. 
In the seminal work by Bistrizer and MacDonald (Ref.~\onlinecite{bistritzer_pnas_2011}), the authors took $u_0=u_1=110~{\rm meV}$, while the authors of Ref.~\onlinecite{koshino_prx_2018} took $u_0 = 79.7~{\rm meV}$ and $u_1 = 97.5~{\rm meV}$. Extensive density functional theory simulations including lattice relaxation~\cite{lucignano_prb_2019, cantele_arXiv_2020} suggest $u_0 = 78~{\rm meV}$ and $u_1 = 98~{\rm meV}$ for a range of twist angles $ 1.08\degree \leq \theta \leq 3.89\degree$.
Finally, the authors of Ref.~\onlinecite{tarnopolsky_prl_2019} considered a greatly simplified continuum model for TBG, which has $u_0=0$. 
In this so-called ``chirally-symmetric" continuum model, the low-energy bands near the CNP are rigorously flat (i.e.~they have zero bandwidth) at the magic angle. As we show below, the values of $u_{0}$ and $u_{1}$ strongly influence the optical properties of TBG. Optical experiments are therefore a very useful tool to measure these parameters.

In this Article we present a thorough investigation of two main physical quantities, namely the {\it local} optical conductivity $\sigma(\omega)$ and energy loss function ${\cal L}({\bm q}, \omega)$ of TBG for different filling factors and twist angles. 
We work at temperatures $T> T_{\rm c}$, where $T_{\rm c}$ is the maximum critical temperature at which any of the aforementioned exotic phases occurs~\cite{cao_nature_2018,cao_nature_2018a,yankowitz_science_2019,sharpe_science_2019,kerelsky_nature_2019,xie_nature_2019,lu_nature_2019,tomarken_prl_2019,jiang_nature_2019,choi_natphys_2019,serlin_arxiv_2019,stepanov_arxiv_2019}. Our theory is believed to be accurate also when e-e interactions are weak enough, i.e.~at low temperatures, provided that $\theta$ is not precisely the magic angle, or, at $\theta=\theta^\star$, provided that the filling factor $|\xi|$ is larger than one---see Eq.~(\ref{eq:nu}) below.
The calculation of optical properties and collective modes of broken symmetry states~\cite{bascones_arxiv_2019} is well beyond the scope of the present Article and is deferred to future publications. 

The optical conductivity is a {\it proper} linear response function relating the electrical current to the {\it total} electric field (i.e.~the sum of the external electric field and the average electric field generated by the electron themselves) applied to an electron system.
It encodes the response of the electron system to a spatially-uniform oscillating field and is therefore of primary importance to interpret far-field optical experiments.
When e-e interactions are neglected, it can be calculated as a sum over allowed transitions according to Kubo formula~\cite{Giuliani_and_Vignale}, once the single-particle eigenstates and eigenenergies of the system are known. Interactions modify this result in essentially two ways.
First, they modify the set of eigenstates that one should use. Indeed, one should add to the non-interacting Hamiltonian a mean-field potential that takes into account the impact on one electron of the presence of all the other electrons.
Second, they add new contributions to the response function stemming from {\it dynamical} exchange and correlation effects.
In the framework of many-body diagrammatic perturbation theory~\cite{Giuliani_and_Vignale}, these can be viewed as arising from {\it irreducible} diagrams containing at least one interaction line.
(Since $\sigma(\omega)$ is a proper response function reducible diagrams do not contribute to the perturbative series.)

With reference to the mean-field theory of linear response (see, for example, Sect. 4.7 of Ref.~\onlinecite{Giuliani_and_Vignale}), we perform our calculations at the level of time-dependent Hartree theory, commonly known, for historical reasons, as Random Phase Approximation (RPA). 
This requires to calculate eigenstates and eigenenergies according to the self-consistent Hartree mean-field theory and then feed these results to the Kubo formula.
The next degree of approximation is the time-dependent Hartree-Fock (TDHF) approximation, which requires the calculations of the Hartree-Fock orbitals and the evaluation of the contributions arising from proper diagrams containing one interaction line. 
Results of the TDHF approximation as applied to TBG will be the subject of a forthcoming publication.

The energy loss function ${\cal L}({\bm q}, \omega)$ measures the amount of energy that the system is able to absorb from an external scalar perturbation with wave vector $\bm q$ and angular frequency $\omega$.
It is particularly useful to identify collective modes that couple to the charge density, since these appear as well defined peaks in the energy loss function.
As explained in Sect.~\ref{sec:optical_cond}, its calculation requires, in principle, the knowledge of the {\it non-local} conductivity $\sigma^{\bm G \bm G'}_{\alpha \beta}(\bm q,\omega)$.
In this work we resort to the local approximation that is appropriate for wavelengths much larger than the moir\'e periodicity and only requires the knowledge of the {\it local} conductivity $\sigma_{\alpha\beta}(\omega)\equiv \lim_{{\bm q} \to {\bm 0}} \sigma^{\bm 0\bm 0}_{\alpha \beta}(\bm q,\omega)$.

As a natural byproduct of the calculation of the intra-band contribution to $\sigma(\omega)$, we obtain an approximate expression for the Seebeck coefficient (or thermopower) $S$.
The latter measures the coupling between electrical and thermal phenomena in TBG.
Our approximation captures the band structure contribution to $S$, while neglecting the largely unknown energy dependence of the scattering mechanisms in TBG.

In this Article we focus on the impact of {\it two} key physical effects on $\sigma(\omega)$ and $\mathcal{L}({\bm q},\omega)$ for TBG with varying filling factor:
\begin{itemize}
\item[i)] As discussed above, recent experiments have highlighted the fact that $u_0$ needs not to be equal or comparable to $u_1$. 
In this work, we fix $u_1$, and study the role of $u_0$ in the range $0\leq u_0 \leq u_1$;

\item[ii)] It has been emphasized~\cite{guinea_pnas_2018,xie_arxiv_2018} that {\it intrinsic} (i.e.~not due to e.g. Coulomb impurities) spatial inhomogeneities are important in magic-angle TBG. 
In other words, due to the moir\'e periodicity, the ground-state electron density $n({\bm r})$ is not homogeneous in space. 
Technically speaking, therefore, the results of the single-particle band models introduced e.g.~in  Refs.~\onlinecite{bistritzer_pnas_2011,koshino_prx_2018} need to be iterated self-consistently in the Hartree approximation~\cite{Giuliani_and_Vignale} to see how they are altered by e-e interactions, as functions of the filling factor. 
Here, we therefore compute the corresponding ``Hartree conductivity" by using the self-consistently calculated bands and eigenstates for a range of filling factors and different twist angles. 
These calculations take into account the role of static screening in reshaping the bare bands and rearranging in space the single-particle Bloch eigenstates of electrons moving in TBG.
\end{itemize}

Moreover, starting from the magic angle, we will study the role of the twist angle $\theta$ in the window $\theta^{\star} \lesssim \theta \lesssim 2\degree$, where the physics of TBG is dominated by the moir\'e modulations of the inter-layer tunneling amplitudes. This regime (and not only the regime $\theta \approx \theta^\star$) is interesting in its own right~\cite{moon_prb_2013} since it displays a markedly different behavior with respect to standard single-layer graphene.

We show that, at zero filling, the optical conductivity and loss function are strongly dependent on $\theta$ and $u_0$, while they are insensitive to Hartree self-consistency. 
Conversely, away from zero filling, the Hartree potential gives strong corrections to both optical conductivity and loss function, especially in the low-frequency domain.

We hasten to emphasize that the optical and plasmonic properties of TBG have been investigated in a number of previous pioneering works~\cite{tabert_prb_2013,moon_prb_2013,stauber_njp_2013,ikeda_arXiv_2020, stauber_nanolett_2016, lewandowski_pnas_2019}. 
The local optical conductivity has been previously calculated by the authors of Refs.~\onlinecite{tabert_prb_2013,moon_prb_2013,stauber_njp_2013} for twist angles $\theta \gtrsim 1.5\degree$. 
These calculations have been carried out by means of the non-interacting continuum model introduced in Ref.~\onlinecite{bistritzer_pnas_2011}. (The optical response of TBG beyond the linear-response approximation has been recently calculated in Ref.~\onlinecite{ikeda_arXiv_2020} for large twist angles.) The loss function of TBG has been calculated in Refs.~\onlinecite{stauber_njp_2013, stauber_nanolett_2016, lewandowski_pnas_2019}, for angles near the magic one, and by means of the non-interacting band models introduced in  Refs.~\onlinecite{bistritzer_pnas_2011,koshino_prx_2018}. 
These works did not take into account neither Hartree self-consistency nor the inter-layer tunnelling asymmetry $u_0\neq u_1$.

Our Article is organized as following. In Sect.~\ref{sec:model} we review the continuum model we have used in this work and briefly summarize the main steps that are needed to incorporate Hartree self-consistency into the theory. 
In Sect.~\ref{sec:optical_cond} we describe in detail the procedure we have used to calculate the optical conductivity and the loss function, briefly commenting on two important byproducts of the general theory, i.e.~the Drude weight and Seebeck coefficient. 
Our main numerical results are presented in Sect.~\ref{sec:numerical_results}. 
A summary of our main findings and a brief set of conclusions are reported in Sect.~\ref{sec:conclusions}. Finally, a number of useful technical details is contained in Appendices~\ref{app:derivation_continuum_model}-\ref{app:technical}.
\section{Electronic band structure}
\label{sec:model}
\begin{figure}[t]
\includegraphics[width = \columnwidth]{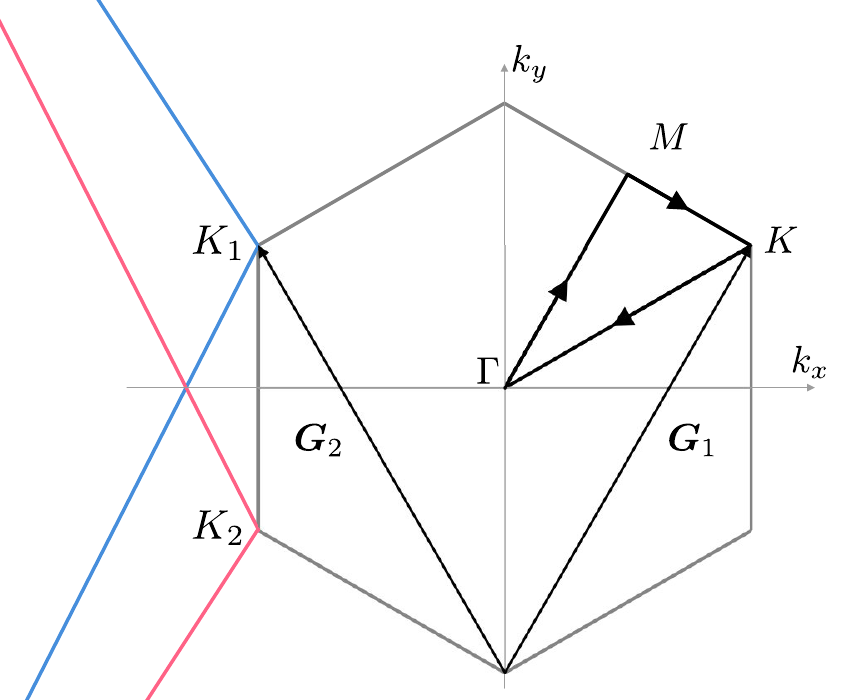}
\caption{\label{fig:mini_bz} (Color online) Moir\'e Brillouin zone of twisted bilayer graphene. The continuum model is a ${\bm k} \cdot {\bm p}$ expansion around the $K_{\ell}$ valley of layer $\ell$. The blue (red) lines are edges of the Brillouin zone of layer 1 (2). The energy band structures shown in Figs.~\ref{fig:bands_var_filling} and \ref{fig:bands_var_delta} are computed along the $K$-$\Gamma$-$M$-$K$ path highlighted here.}
\end{figure}

The continuum description of electrons roaming in TBG adopted in this work was first derived in Refs.~\onlinecite{bistritzer_pnas_2011,koshino_prx_2018}. We refer the reader to these works and Appendix~\ref{app:derivation_continuum_model} for all relevant technical details.

Layer, sublattice, spin, and valley are the four discrete degrees of freedom characterizing single-electron states in TBG. 
As shown in Refs.~\onlinecite{bistritzer_pnas_2011,koshino_prx_2018,guinea_pnas_2018}, valley and spin degrees of freedom are accounted for by a fourfold degeneracy factor. The single-particle Hamiltonian of TBG is therefore written in the layer/sublattice basis $\{|1A\rangle,|1B\rangle,|2A\rangle,|2B\rangle\}$, where it takes the form
\begin{equation}\label{eq:abstract_hamiltonian}
 	\hat{H}_{0} = \begin{pmatrix}\hat{H}^{(1)} && \hat{U}\\  \hat{U}^{\dagger}&& \hat{H}^{(2)}\end{pmatrix}~.
\end{equation}
The state $|\ell\tau\rangle$ refers to layer $\ell=1,2$ and sublattice index $\tau=A,B$, $\hat{H}^{(\ell)}$ is the intra-layer Hamiltonian for layer $\ell$, and the operator $\hat{U}$ describes inter-layer tunneling.
For small twist angles, the moir\'e length scale $a/\theta$ is much larger than the lattice parameter of monolayer graphene.
This allows us to safely replace $\hat{H}^{(\ell)}$ by its ${\bm k} \cdot {\bm p}$ expansion, i.e.~by the following massless Dirac fermion Hamiltonian: 
\begin{equation}\label{eq:intralayer_hamiltonian}
 	\hat{H}^{(\ell)} = v_{\rm{D}}\left[R_\ell(\theta/2)(\hat{\bm{p}} - \hbar\bm{K}_{\ell})\right]\cdot(\sigma_{x},\sigma_{y})~.
\end{equation}
Here, $(\sigma_{x},\sigma_{y})$ is a vector of ordinary $2\times 2$ Pauli matrices, $\hat{\bm{p}}$ is the momentum operator, $v_{{\rm D}}=\sqrt{3}|t|a/(2\hbar) \simeq 0.87 \times 10^{6} {\rm m}/{\rm s}$ is the Fermi velocity of monolayer graphene---corresponding to the standard nearest-neighbor hopping energy of $|t|=2.7 \, {\rm eV}$ adopted in tight-binding models of graphene---and $\bm{K}_{\ell}$ is the position of graphene's valley $K_{\ell}$ measured from the MBZ center $\Gamma$ (see Fig.~\ref{fig:mini_bz}), i.e.
\begin{equation}\label{eq:monolayer_valleys}
\bm{K}_{1, 2} = \frac{8\pi}{3a}\sin \left(\frac{\theta}{2}\right)\left(-\frac{\sqrt{3}}{2}, \pm\frac{1}{2}\right)~.
\end{equation}
Finally, the $2 \times 2$ rotation matrix
\begin{equation}\label{eq:rotation_matrix}
R_{1, 2}(\theta/2) = \cos (\mp \theta/2) \openone_{2\times 2} -i\sin(\mp \theta/2) \sigma_{y}~,
\end{equation}
$\openone_{2\times 2}$ being the identity matrix, accounts for the orientation of graphene's original Brillouin zones.
In these formulas, the upper (lower) sign applies to $\ell =1$ ($\ell =2$).

The operator $\hat{U}$ appearing in Eq.~\eqref{eq:abstract_hamiltonian} is the inter-layer tunneling term, given by
\begin{equation}\label{eq:inter_layer_hamiltonian}
\begin{split}
\hat{U} = 
\begin{pmatrix}
u_0 && u_1 \\
u_1 && u_0
\end{pmatrix} 
+  e^{-i\frac{2\pi}{3} + i\bm{G}_{1}\cdot\hat{\bm{r}}} &
\begin{pmatrix}
u_0 && u_1e^{i\frac{2\pi}{3}} \\
u_1e^{-i\frac{2\pi}{3}}&& u_0
\end{pmatrix}
+ \\ 
+ e^{i\frac{2\pi}{3} + i\bm{G}_{2}\cdot\hat{\bm{r}}} &
\begin{pmatrix}
u_0 && u_1e^{-i\frac{2\pi}{3}} \\
u_1e^{i\frac{2\pi}{3}} && u_0
\end{pmatrix}
~,
\end{split}	
\end{equation}
where $u_1$ and $u_0$ are the  inter- and intra-sublattice hopping energies, respectively, and  $\{\bm{G}_{1}, \bm{G}_{2}\}$ are primitive vectors of the moir\'e reciprocal lattice,
\begin{equation}\label{eqs:reciprocal_lattice_vectors}
\bm{G}_{1, 2} = \frac{8\pi}{a\sqrt{3}}\sin \left(\frac{\theta}{2}\right)\left(\pm\frac{1}{2},\frac{\sqrt{3}}{2}\right)~, 
\end{equation}
as illustrated in Fig.~\ref{fig:mini_bz}.

In Ref.~\onlinecite{bistritzer_pnas_2011} the inter- and intra-sublattice hopping energies were fixed to the same value, $u_1 = u_0 = 110~{\rm meV}$, known from the theory of aligned (i.e.~Bernal stacked) bilayer graphene. In more recent publications~\cite{koshino_prx_2018,carr_prresearch_2019,lucignano_prb_2019}, it was noticed that a difference between $u_0$ and $u_1$ can effectively account for lattice corrugations occurring in TBG sheets. 
Moreover, in a recent experimental work~\cite{hesp_arxiv_2019}, the measured inter-band collective excitations in TBG have been well reproduced by a model with a much smaller intra-sublattice hopping energy $u_0$ than the one in Ref.~\onlinecite{koshino_prx_2018}. 
Inter-sublattice and intra-sublattice hopping energies, indeed, may also be affected by extrinsic factors such as strain originated in the sample fabrication procedure. 

Given these facts, in this work we consider the intra-sublattice hopping energy $u_0$ as a free parameter of the model, ranging from $u_0 = 0$ as in Ref.~\onlinecite{tarnopolsky_prl_2019} to $u_0 = u_1$ as in Ref.~\onlinecite{bistritzer_pnas_2011}. For a fixed value of the inter-sublattice hopping energy $u_1$, the magic angle is weakly dependent on $u_0$ (see Ref.~\onlinecite{tarnopolsky_prl_2019}).
In this work, we set $u_1 = 97.5~{\rm meV}$, as in Ref.~\onlinecite{koshino_prx_2018}, which yields a magic angle $\theta^{\star} \simeq 1\degree$, slightly varying with $u_0$.

\subsection{Self-consistent Hartree theory}
\label{sec:Hartree}

The single-particle problem posed by Eqs.~(\ref{eq:abstract_hamiltonian})-(\ref{eqs:reciprocal_lattice_vectors}) can be easily solved numerically and one can find energy bands and corresponding Bloch states~\cite{bistritzer_pnas_2011,koshino_prx_2018}.

It is, however, well known~\cite{guinea_pnas_2018,koshino_prx_2018} that the charge density distribution in TBG is fairly inhomogeneous. And such inhomogeneities create an inhomogeneous electric potential that depends on the filling factor. 
To incorporate this effect into the theory, we need to diagonalize the following Hamiltonian 
\begin{equation}\label{eq:abstract_full_hamiltonian}
\hat{H} = \hat{H}_{0} + \hat{V}_{\rm H}[n_{\bm{G}}]~,
\end{equation}
where
\begin{equation}\label{eq:hartree_potential}
\hat{V}_{\rm H}[n_{\bm{G}}] = \openone_{4\times 4}\sum_{\bm{G} \neq \bm{0}}\frac{2\pi e^{2}}{\bar{\varepsilon}(0)|\bm{G}|}n_{\bm{G}}e^{i\bm{G}\cdot\hat{\bm{r}}}~,
\end{equation}
is the Hartree potential~\cite{Giuliani_and_Vignale} and $\bar{\varepsilon}(0)$ is a suitable static dielectric function (see Sect.~\ref{sect:energy_loss_function}).
Here, $\openone_{4\times 4}$ is the $4\times4$ identity matrix in the basis $\{|1A\rangle,|1B\rangle,|2A\rangle,|2B\rangle\}$, the sum runs over the non-zero moir\'e reciprocal lattice vectors $\bm{G}$ (the term with ${\bm G}={\bm 0}$ is excluded to ensure overall charge neutrality due to the positively charged background~\cite{Giuliani_and_Vignale}), whereas $n_{\bm{G}}$ is the Fourier component of the ground-state electron density corresponding to the reciprocal lattice vector $\bm G$. 
The problem posed by Eqs.~(\ref{eq:abstract_full_hamiltonian})-(\ref{eq:hartree_potential}) needs to be solved self-consistently, i.e.~one needs to solve the Hartree equation
\begin{equation}\label{eq:self_consistency:hamiltonian}
 \left(\hat{H}_{0} + \hat{V}_{\rm{H}}[n_{\bm{G}}] \right)|\bm{k}\nu\rangle = \epsilon_{\bm{k}\nu} |\bm{k}\nu\rangle~, 
\end{equation}
together with the self-consistency closure
\begin{equation}\label{eq:self_consistency:density}
 n_{\bm{G}} = \frac{g}{A}
 \sum_{\bm{k}, \nu}f_{\bm{k}\nu}\langle\bm{k}\nu|e^{-i\bm{G}\cdot\hat{\bm{r}}}|\bm{k}\nu\rangle~.
\end{equation}
Here, the factor $g = 4$ accounts for the spin/valley degeneracy, $A$ is the 2D electron system area, $\nu$ is a band index, $f_{\bm{k}\nu} = \{1+ \exp[(\epsilon_{\bm{k}\nu}-\mu)/(k_{\rm B}T)]\}^{-1}$ is the Fermi-Dirac distribution function, and $\mu$ is the chemical potential. The derivation of Eq.~\eqref{eq:hartree_potential} is reported in Appendix~\ref{app:hartree_potential_derivation}. 

At a given  temperature $T$, the chemical potential can be found by the usual equation ensuring particle-number conservation:
\begin{equation}\label{eq:self_consistency:average_density}
\delta n + n_{0} = g \sum_{\nu} \int_{\rm MBZ} \frac{d^2{\bm k}}{(2\pi)^2}f_{\bm{k}\nu}~.
\end{equation}
Here, the electron (hole) density $\delta n>0$ ($\delta n<0$) is simply the electron density measured from the CNP, i.e.~$\delta n = 0$ at CNP, and the quantity $n_{0}$ is the total electron density at CNP. 
The latter can be conveniently expressed in units of the following ``elementary density", corresponding to the contribution to the total density coming from a fully occupied energy band:
\begin{equation}\label{eq:density_1band}
n_{\rm b} = \frac{g}{\Omega_{{\rm u.c.}}}~,
\end{equation}
where
\begin{equation}
\Omega_{\rm{u.c.}} = \frac{\sqrt{3}}{2}\left[\frac{a}{2\sin(\theta/2)}\right]^{2}
\end{equation}
is the area of the moir\'e unit cell.
The low-energy continuum model predicts the existence of an infinite number of moir\'e mini-bands above and below the CNP.
If we retain a number $N_{{\rm bands}}$ of energy bands above the CNP and $N_{{\rm bands}}$ energy bands below the CNP, the density at the CNP is $n_0 = N_{{\rm bands}}n_{{\rm b}}$. 
The ``filling factor" is therefore given by the dimensionless ratio:
\begin{equation}\label{eq:nu}
\xi \equiv \frac{\delta n}{n_{\rm b}}~.
\end{equation}
At zero filling, $\delta n = 0$ and $\mu\approx 0$ ($\mu$ is not exactly zero at zero filling because particle-hole symmetry is not exact). 
In particular, the chemical potential is within the flat bands when $|\xi| < 1$ and temperature is small.

For completeness we mention that an equivalent definition of the filling factor in which the degeneracy factor $g$ is not included in Eq.~\eqref{eq:density_1band} is also commonly found in literature. In that case the chemical potential of TBG is within the flat bands when $|\xi| < 4$ (and temperature is small).

Because the denominator in Eq.~\eqref{eq:hartree_potential} grows with $|\bm{G}|$, in solving the equations self-consistently we can limit the sum over $\bm{G}$ to the first hexagonal shell spanned by the primitive vectors in Eq.~\eqref{eqs:reciprocal_lattice_vectors}. Contributions to the sum coming from outer shells with larger values of $|\bm{G}|$ are strongly suppressed~\cite{guinea_pnas_2018}. The convergence of the self-consistent procedure delicately depends on the number of bands used in the summation in Eq.~(\ref{eq:self_consistency:density}). This and other important numerical details are reported in Appendix~\ref{app:technical}.

In writing the eigenvalue equation \eqref{eq:self_consistency:hamiltonian} and the self-consistency closure (\ref{eq:self_consistency:density}), we have implicitly assumed that e-e interactions do not break the discrete translational symmetry of the original problem defined by $\hat{H}_{0}$. We have therefore chosen the eigenstates to be of the Bloch form
\begin{equation}
|\bm{k}\nu\rangle = e^{i\bm{k}\cdot\bm{r}}|u_{\bm{k}\nu}\rangle~,
\end{equation}
where $|u_{\bm{k}\nu}\rangle$ is the periodic part of the Bloch function. 
For future reference, we introduce the $\bm k$-dependent Hamiltonian
\begin{equation}\label{eq:acca_kappa}
\begin{split}
\hat{H}(\bm k) &\equiv e^{-i\bm k \cdot \hat{\bm r}}\left(\hat{H}_{0} + \hat{V}_{\rm{H}}[n_{\bm{G}}]\right)e^{i\bm k \cdot \hat{\bm r}}\\ & =\begin{pmatrix} \hat{H}^{(1)}(\bm k) && \hat{U}\\ \hat{U}^\dagger && \hat{H}^{(2)}(\bm k)\end{pmatrix}~  + \hat{V}_{\rm{H}}[n_{\bm{G}}],
\end{split}
\end{equation}
\vspace{15mm}
where $\hat{H}^{(\ell)}(\bm k)\equiv e^{-i\bm k \cdot \hat{\bm r}}\hat{H}^{(\ell)}e^{i\bm k \cdot \hat{\bm r}}$ are easily obtained from  $\hat{H}^{(\ell)}$ by replacing $\hat{\bm p}$ with $\hat{\bm p} + \hbar \bm k$ in Eq.~(\ref{eq:intralayer_hamiltonian}).

\section{Optical conductivity, Drude weight, Seebeck coefficient, and energy loss function}
\label{sec:optical_cond}
In this Section, we define the key quantities we have calculated in this work, i.e.~the optical conductivity $\sigma(\omega)$, the Drude weight ${\cal D}$, the Seebeck coefficient $S$, and the energy loss function ${\cal L}({\bm q}, \omega)$.

\subsection{Optical conductivity and Drude weight}
\label{sect:optical}
The optical conductivity $\sigma_{\alpha\beta}(\omega)$ is the linear-response function relating the electrical current flowing in the direction $\alpha$ in response to the {\it total} electric field applied in the direction $\beta$.
In crystals it can be separated into an intra-band and an inter-band contribution,
\begin{equation}\label{eq:optical_conductivity}
  \sigma_{\alpha\beta}(\omega) = \sigma_{\alpha\beta}^{\rm{intra}}(\omega) + \sigma_{\alpha\beta}^{\rm{inter}}(\omega)~.
\end{equation}
Both contributions can be calculated by using the Kubo formula (see Appendix~\ref{app:conductivity}).
The intra-band contribution has a simple Drude-type frequency dependence and is given by
\begin{equation}\label{eq:optical_conductivity_intra}
\sigma_{\alpha\beta}^{\rm{intra}}(\omega) = G_0 \frac{i\mathcal{W}_{\alpha\beta}^{(0)}}{\hbar\omega + i\eta}~,
\end{equation}
where $G_0 \equiv 2e^{2}/h$ is the conductance quantum, $\eta$ is a small positive infinitesimal (with dimensions of energy) and $\mathcal{W}_{\alpha\beta}^{(0)}$ can be calculated from
\begin{widetext}
\begin{equation}\label{eq:drude_weight}
\mathcal{W}_{\alpha\beta}^{(p)} \equiv -\pi g\sum_{\nu}\int \frac{d^2\bm{k}}{(2\pi)^{2}}f_{\bm{k}\nu}^{\prime} (\epsilon_{\bm{k}\nu}-\mu)^p \langle u_{\bm{k}\nu} |\partial_{k_{\alpha}}\hat{H}(\bm{k})|u_{\bm{k}\nu}\rangle\langle u_{ \bm{k}\nu }|\partial_{k_{\beta}}\hat{H}(\bm{k})|u_{\bm{k}\nu}\rangle~,
 \end{equation}
\end{widetext}
by setting $p=0$. 
In Eq.~(\ref{eq:drude_weight}), the factor $(\epsilon_{\bm{k}\nu}-\mu)^p$ with $p$ a non-negative integer has been introduced for later convenience, $f_{\bm{k}\nu}^{\prime}$ denotes the derivative of the Fermi distribution with respect to its argument, and the factor $g = 4$ accounts for the aforementioned fourfold valley/spin degeneracy. 
The quantity $\mathcal{W}_{\alpha\beta}^{(0)}$ is proportional to the Drude weight $\mathcal{D}_{\alpha\beta}$, i.e.
\begin{equation}
\mathcal{D}_{\alpha\beta} \equiv \int_{-\infty}^{\infty}d\omega \re [\sigma_{\alpha\beta}^{\rm{intra}}(\omega)] = \frac{e^2}{\hbar^2}\mathcal{W}_{\alpha\beta}^{(0)}~.
\end{equation}
The inter-band contribution to the optical conductivity is given by
\begin{widetext}
\begin{equation}\label{eq:optical_conductivity_inter}
\sigma_{\alpha\beta}^{\rm{inter}}(\omega) = -i\pi  g G_0 \sum_{\nu \neq \nu^{\prime}}\int \frac{d^2\bm{k}}{(2\pi)^{2}} \frac{f_{\bm{k}\nu} - f_{\bm{k}\nu^{\prime}}}{\epsilon_{\bm{k}\nu} - \epsilon_{\bm{k}\nu^{\prime}}}
\frac{\langle u_{\bm{k}\nu}|\partial_{k_{\alpha}}\hat{H}(\bm{k})|u_{\bm{k}\nu^{\prime}}\rangle\langle u_{\bm{k}\nu^{\prime}} |\partial_{k_{\beta}}\hat{H}(\bm{k})|u_{\bm{k}\nu}\rangle}{\hbar\omega +i\eta + \epsilon_{\bm{k}\nu} - \epsilon_{\bm{k}\nu^{\prime}}}~.
\end{equation}
\end{widetext}
For 2D systems, the optical conductivity is in general a $2\times 2$ matrix with respect to the Cartesian indices $\alpha,\beta$. 
Since the Hamiltonian in Eq.~\eqref{eq:abstract_hamiltonian} has a $D_{3}$ point group~\cite{koshino_prx_2018}, which we assume to be unbroken also when e-e interactions are taken into account in the Hartree approximation (\ref{eq:abstract_full_hamiltonian}), it follows that $\sigma_{\alpha\beta}(\omega) = \sigma(\omega)\delta_{\alpha\beta}$, where $\sigma(\omega) \equiv \sigma_{xx}(\omega) =\sigma_{yy}(\omega)$ and $\delta_{\alpha\beta}$ is the Kronecker symbol.
The same holds for all the other relevant properties, i.e. $\sigma^{{\rm intra}}_{\alpha\beta}(\omega) = \sigma^{{\rm intra}}(\omega)\delta_{\alpha\beta}$, $\sigma^{{\rm inter}}_{\alpha\beta}(\omega) = \sigma^{{\rm inter}}(\omega)\delta_{\alpha\beta}$,$\mathcal{W}^{(p)}_{\alpha\beta} = \mathcal{W}^{(p)}\delta_{\alpha\beta}$ and $\mathcal{D}_{\alpha\beta} = \mathcal{D}\delta_{\alpha\beta}$.

\subsection{Seebeck coefficient in the relaxation time approximation}

Integrals of the type written in Eq.~(\ref{eq:drude_weight}) are also useful to calculate the Seebeck coefficient $S$, which describes the electrical response to a thermal gradient. 

Indeed, the Seebeck coefficient can be written as~\cite{Ashcroft_and_Mermin}
\begin{equation}\label{eq:seebeck_full}
S = -\frac{1}{eT} \frac{I^{(1)}}{I^{(0)}}~,
\end{equation}
where
\begin{eqnarray}\label{eq:I_def}
I^{(p)}&\equiv&  -\frac{e^2}{\hbar^2} g \sum_{\nu}\int \frac{d^2\bm k}{(2\pi)^2} f_{\bm{k}\nu}^{\prime} \tau_{\bm k\nu}(\epsilon_{\bm{k}\nu}-\mu)^p \nonumber\\
&\times&
\langle u_{\bm{k}\nu} |\partial_{k_{x}}\hat{H}(\bm{k})|u_{\bm{k}\nu}\rangle\langle u_{ \bm{k}\nu }|\partial_{k_{x}}\hat{H}(\bm{k})|u_{\bm{k}\nu}\rangle~.
\end{eqnarray}
Here, $\tau_{\bm k\nu}$ is the momentum-dependent relaxation time. 
In the Relaxation Time Approximation (RTA), where the dependence of $\tau_{{\bm k}\nu}$ on ${\bm k}$ is neglected by setting $\tau_{{\bm k}\nu}\equiv \tau$, Eq.~(\ref{eq:seebeck_full}) reduces to
\begin{equation}\label{eq:seebeck_approx}
S_{\rm RTA} = -\frac{1}{eT}\frac{\mathcal{W}^{(1)}}{\mathcal{W}^{(0)}}~.
\end{equation}
The RTA neglects the energy and momentum dependence of the scattering time, but correctly captures the intrinsic (i.e.~band structure) contribution to the Seebeck coefficient.

\subsection{Energy loss function and plasmons}
\label{sect:energy_loss_function}
The energy loss function (or, briefly, loss function) $\mathcal{L}(\bm{q},\omega)$ is proportional to the probability of exciting the 2D electron system by applying a scalar perturbation of wave vector $\bm{q}$ and energy $\hbar\omega$. 
The loss function can be directly measured e.g.~via electron-energy-loss spectroscopy~\cite{egerton_rep_prog_phys_2009} and displays peaks where self-sustained charge oscillations--- i.e.~plasmons---can be excited. 
It also carries information on inter-band transitions and Landau damping. 
As mentioned in Sect.~\ref{sect:intro}, collective excitations of 2D electron systems can also be probed by scattering-type near-field optical microscopy. 
We refer the reader to Ref.~\onlinecite{hesp_arxiv_2019} for results of this experimental technique as applied to TBG.

In a crystal, the loss function is formally defined by~\cite{tomadin_prb_2014}
\begin{equation}\label{eq:loss_function_definition}
\mathcal{L}(\bm{q},\omega) = -\im \left\{[\epsilon^{-1}(\bm{q},\omega)]_{\bm G = \bm 0,\bm G'=\bm 0}\right\}~.
\end{equation}
Here, $\epsilon_{\bm G\bm G'}(\bm{q},\omega)$ is the dielectric function of the crystal~\cite{Giuliani_and_Vignale} viewed as a matrix with indices $\bm G$, $\bm G'$ in the space of reciprocal lattice vectors, $\bm q$ lies inside the first Brillouin zone, and inversion has to be understood as {\it matrix} inversion.

The dielectric function can in turn be expressed as
\begin{equation}\label{eq:non_local_epsilon}
\begin{split}
&\epsilon_{\bm G\bm G'}(\bm{q},\omega) =\\
&=\delta_{\bm G\bm G'} +   L_{\bm q+\bm G, \omega}\frac{i(\bm q + \bm G)_\alpha(\bm q + \bm G')_\beta \sigma_{\alpha\beta}^{\bm G\bm G'}(\bm q,\omega)}{\omega}~.
\end{split}
\end{equation}
Here, $L_{\bm{q}+{\bm G}, \omega}$ is the Coulomb interaction potential relating charge density fluctuations $\rho(\bm{q}+{\bm G}, \omega)$ to the self-induced electrical potential $\phi(\bm{q}+{\bm G},\omega)$, i.e. $\phi(\bm{q}+{\bm G},\omega)=L_{\bm{q}+{\bm G}, \omega}\rho(\bm{q}+{\bm G}, \omega)$, and $\sigma_{\alpha\beta}^{\bm G\bm G'}(\bm q,\omega)$ is the non-local conductivity.
We refer the reader to Appendix~\ref{app:non_local_dielectric_function} and references cited therein for details on the derivation of Eq.~(\ref{eq:non_local_epsilon}).

In the following we calculate the loss function in the local approximation. This amounts to neglecting the off-diagonal ${\bm G}\neq {\bm G}^\prime$ terms in the space of the reciprocal lattice vectors and taking the limit ${\bm q} \to {\bm 0}$ in the non-local conductivity, i.e. 
\begin{equation}\label{eq:local_epsilon}
\begin{split}
&\epsilon_{\bm G\bm G'}(\bm{q},\omega) \approx \delta_{\bm G\bm G'}\Bigg[1 + L_{\bm q+\bm G, \omega}  \times \\
&\times \frac{i(\bm q + \bm G)_\alpha(\bm q + \bm G)_\beta \lim_{{\bm q} \to {\bm 0}} \sigma_{\alpha\beta}^{\bm G\bm G}(\bm q,\omega)}{\omega}\Bigg]~.
\end{split}
\end{equation}
By following this procedure and making use of the isotropy of the system (discussed in Sect.~\ref{sect:optical}), we can express ${\cal L}({\bm q},\omega)$ solely in terms of the local conductivity $\lim_{{\bm q} \to {\bm 0}} \sigma_{\alpha\beta}^{\bm 0\bm 0}(\bm q,\omega) =\delta_{\alpha\beta}\sigma(\omega)$ and the interaction potential $L_{\bm q,\omega}$:
\begin{equation}
\label{eq:local_loss_function}
{\cal L}(\bm q,\omega) \approx -\im\left\{\frac{1}{\displaystyle 1 + iq^2L_{\bm q,\omega}\frac{\sigma(\omega)}{\omega}}\right\}~.
\end{equation}
In a 2D system sandwiched between two half-spaces filled with a dielectric with a frequency-dependent permittivity $\bar{\varepsilon}(\omega)$, the interaction potential appearing in Eq.~\eqref{eq:local_loss_function} reads as following:
\begin{equation}\label{eq:interaction_potential}
L_{{\bm q},\omega} = \frac{2\pi}{q\bar{\varepsilon}(\omega)}~.
\end{equation}
In the main text of this Article we present  numerical results for the case of a frequency-independent permittivity, i.e.~we set $\bar{\varepsilon}(\omega) = \bar{\varepsilon}(0)~\forall ~\omega$, thereby neglecting extrinsic dynamical screening effects, which change from dielectric material to dielectric material. In Appendix~\ref{app:hBN_loss}, however, we discuss the plasmonic properties of TBG encapsulated between two hexagonal Boron Nitride (hBN) slabs, where the frequency dependence of $\bar{\varepsilon}$ in the mid-infrared spectral range is significant.
\section{Numerical Results}
\label{sec:numerical_results}
In this Section we present our main numerical results. As stated in Sect.~\ref{sec:model}, in this work we set the inter-sublattice hopping energy to $u_1 = 97.5~{\rm meV}$. Most of our calculations below have been carried out at a twist angle $\theta = 1.05\degree$, which is close to the magic angle~\cite{koshino_prx_2018,tarnopolsky_prl_2019,guinea_pnas_2018}. 

Dependencies on the twist angle are presented in Sect.~\ref{subsec:var_theta} below. 

All our numerical results have been obtained by setting $T = 5~{\rm K}$ and $\bar{\varepsilon}(0) = 4.9$. 
\begin{figure*}[t]
\includegraphics[width = \textwidth]{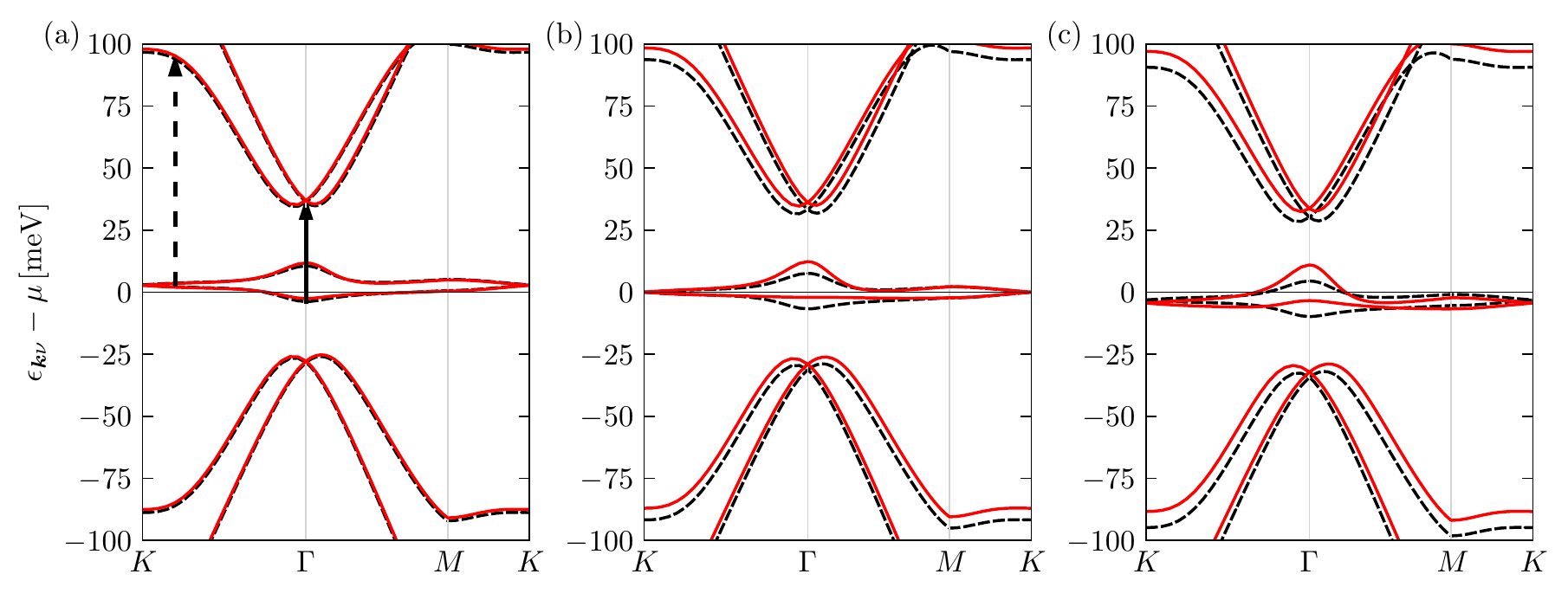}
\caption{\label{fig:bands_var_filling} (Color online) TBG energy bands for different filling factors are plotted along the $K$-$\Gamma$-$M$-$K$ path in the MBZ (see Fig.~\ref{fig:mini_bz}). Data in this figure refer to $\theta = 1.05\degree$, $u_0 = 79.7~{\rm meV}$, $u_1 = 97.5~{\rm meV}$, $T = 5~{\rm K}$, and $\bar{\varepsilon}(0) = 4.9$. Solid red and dashed black lines are the energy bands calculated by including e-e interactions in the Hartree self-consistent approach discussed in Sect.~\ref{sec:Hartree} and with the single-particle Hamiltonian \eqref{eq:abstract_hamiltonian}, respectively. Different panels refer to different values of the filling factor $\xi$. Panel (a) Hole doping: $\xi = - 3/4$. The vertical arrows mark the optical transitions responsible for the peaks in the optical conductivity, as discussed in Sections \ref{subsec:var_filling}, \ref{subsec:var_delta} and \ref{subsec:var_theta}. Panel (b) CNP: $\xi=0$. Panel (c) Electron doping: $\xi = +3/4$.}
\end{figure*}
\begin{figure*}[t]
\includegraphics[width = \textwidth]{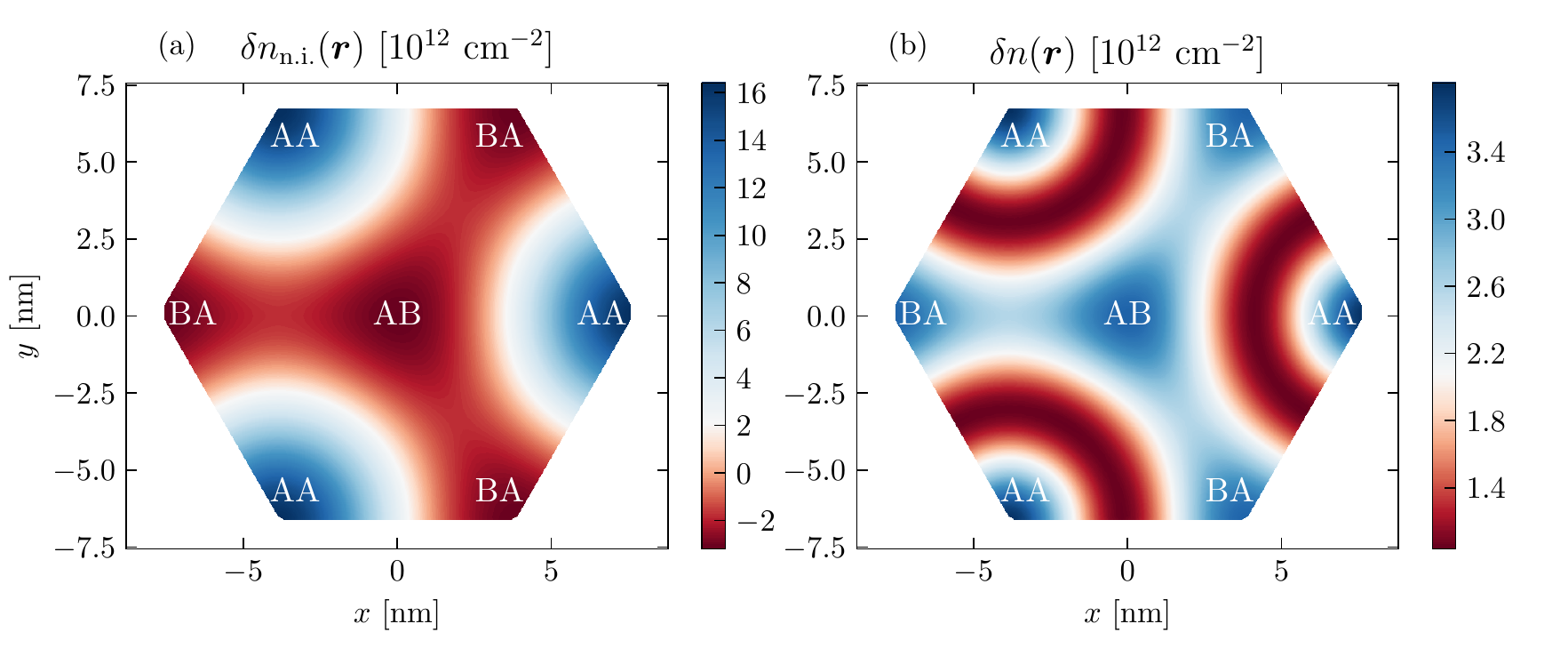}
\caption{\label{fig:density_corrections} (Color online) Deviations $\delta n({\bm r})$ of the ground-state density $n({\bm r})$ from the density $n_0$ at the CNP. The quantity in Eq.~(\ref{eq:real_space_density_deviations}) is plotted for filling factor $\xi = +3/4$. Data in this figure refer to $\theta = 1.05\degree$, $u_0 = 79.7~{\rm meV}$, $u_1 = 97.5~{\rm meV}$, $T = 5~\rm K$, and $\bar{\varepsilon}(0) = 4.9$. Panel~(a) Results for non-interacting electrons. Panel~(b) Results calculated by taking into account the Hartree potential~\eqref{eq:hartree_potential}. In both panels the density value corresponding to the white color is set to $\delta n_{{\rm n.i.}} = \delta n \simeq 2.07\times 10^{12}~{\rm cm}^{-2}$, corresponding to the average value of the density throughout the primitive cell. We have annotated the regions of the unit cell where local AB, AA, and BA stacking occurs between the two layers.}
\end{figure*}
\subsection{Dependence on the filling factor}\label{subsec:var_filling}
In this Section we discuss dependencies of the various quantities introduced in Sect.~\ref{sec:optical_cond} on the filling factor. Here, we set the intra-sublattice hopping energy at the value~\cite{koshino_prx_2018} $u_0 = 79.7~{\rm meV}$.

In Fig.~\ref{fig:bands_var_filling} we plot the moir\'e bands of TBG for three values of the filling factor. In the absence of the Hartree potential, the band structure (black dashed lines) is independent of the filling and is composed of flat bands close to zero energy and higher energy dispersive bands with positive (conduction bands) and negative (valence bands) energy, in agreement with the results of Ref.~\onlinecite{koshino_prx_2018}.
At zero temperature and zero filling ($\mu=0$), the valence flat band and lower-energy valence bands are fully occupied. The conduction flat band and higher-energy bands, on the other hand, are completely empty.

When the Hartree potential is taken into account, all the energy bands (solid red lines in Fig.~\ref{fig:bands_var_filling}) exhibit a filling-factor-dependent distortion with respect to the bare bands. In the corners of the MBZ, i.e.~in the vicinity of the $K$ points, the distortion due to the Hartree potential  is negligible and virtually filling independent, whereas it becomes prominent in the neighbourhood of the MBZ's center, i.e.~the $\Gamma$ point. When $\xi= -3/4$  [see Fig.~\ref{fig:bands_var_filling}(a)], the bands' distortion is moderate throughout the MBZ (this is valid also for higher and lower energy bands).
At zero filling ($\xi=0$) [see Fig.~\ref{fig:bands_var_filling}(b)] and for $\xi=+3/4$ [see Fig.~\ref{fig:bands_var_filling}(c)], however, the flat bands display a substantial upward bending, up a value larger than $\sim 5~{\rm meV}$ at the $\Gamma$ point [see Fig.~\ref{fig:bands_var_filling}(c)].
We point out that such distortion is of the same order of the flat-band bandwidth.
The strong impact of Hartree corrections on the flat bands of TBG was already highlighted in Ref.~\onlinecite{guinea_pnas_2018}.
Higher and lower energy bands are also affected by the Hartree potential by a virtually rigid upward energy shift, with little shape distortion.

In Fig.~\ref{fig:density_corrections} we show how the real space density $n({\bm r})$ deviates from the density $n_{0}$ at the CNP, i.e.~we plot the quantity
\begin{equation}\label{eq:real_space_density_deviations}
\delta n(\bm{r}) \equiv n(\bm{r}) - n_{0}~,
\end{equation}
where $n_{0}$ was defined in Eq.~\eqref{eq:self_consistency:average_density} and
\begin{equation}\label{eq:real_space_density_profile}n(\bm{r}) \equiv  \sum_{\bm{G}} n_{\bm{G}}e^{i\bm{G}\cdot\bm{r}}~.
\end{equation}
The sum over $\bm{G}$ in the Eq.~(\ref{eq:real_space_density_profile}) runs over the vectors in the first hexagonal shell spanned by the primitive vectors in Eq.~\eqref{eqs:reciprocal_lattice_vectors}, whereas $n_{\bm{G}}$ was defined in Eq.~\eqref{eq:self_consistency:density}. For a full derivation of Eq.~\eqref{eq:real_space_density_profile} we refer the reader to Appendix~\ref{app:hartree_potential_derivation}. Numerical results in Fig.~\ref{fig:density_corrections} refer to $\xi = +3/4$. 

In panel (a) of Fig.~\ref{fig:density_corrections} we plot the non-interacting density profile $\delta n_{{\rm n.i.}}(\bm{r})$, which is calculated by neglecting the Hartree potential. It displays spatial fluctuations across the primitive cell on the order of $\lesssim 20 \times 10^{12}~{\rm cm}^{-2}$. On the other hand, when the Hartree potential (i.e.~screening) is taken into account as in panel (b), the amplitude of density oscillations is significantly reduced to $\lesssim 3 \times 10^{12}~{\rm cm}^{-2} $. In the two panels we have set the center of the diverging color map (i.e.~the value corresponding to the white color) to $\delta n_{{\rm n.i.}} = \delta n \simeq 2.07\times 10^{12}~{\rm cm}^{-2}$, corresponding to the average value of the density throughout the primitive cell. Indeed, a simple integration of Eq.~\eqref{eq:real_space_density_deviations} over the unit cell of TBG yields 
\begin{equation}
  \frac{1}{\Omega_{\rm u.c.}}\int_{{\rm u.c.}} d\bm{r} ~\delta n (\bm{r}) = \delta n = \xi n_{{\rm b}},   
\end{equation}  
where $\delta n$ was defined in Eq.~\eqref{eq:self_consistency:average_density} and the last equality follows from Eq.~\eqref{eq:nu}.
\begin{figure*}[t]
  \includegraphics{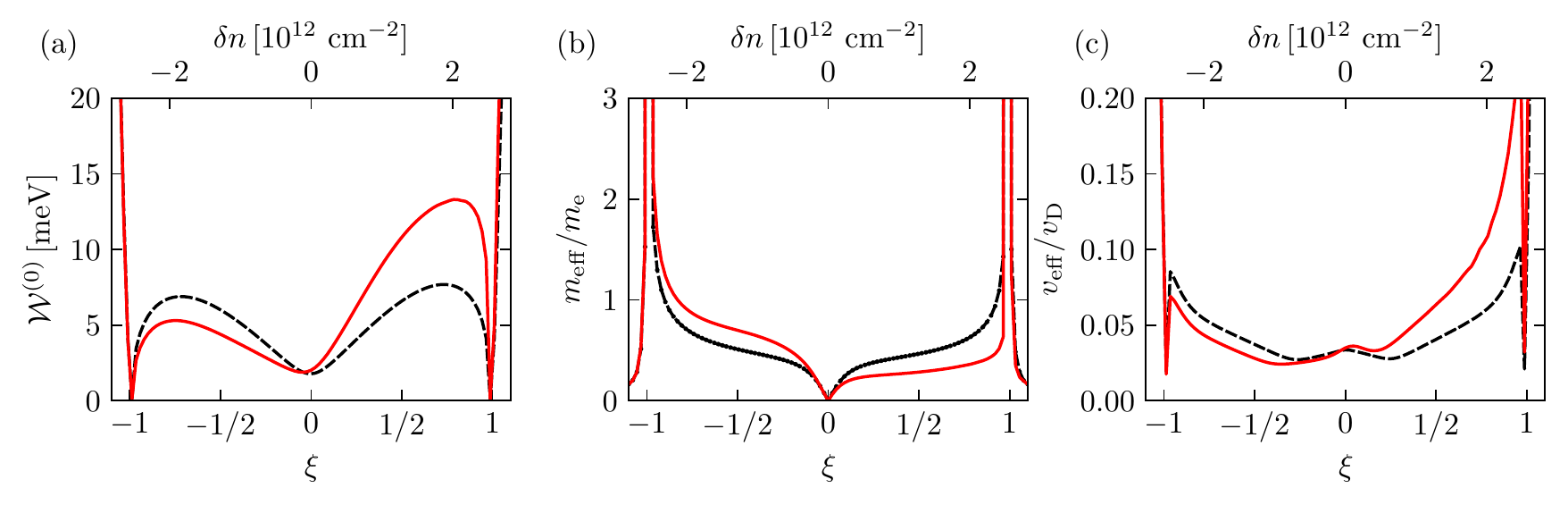}
  \caption{\label{fig:drude_var_filling} (Color online) Drude weight and related physical quantities as functions of the carrier density $\delta n$ (upper horizontal axis) and filling factor $\xi$ (lower horizontal axis), for $\theta = 1.05\degree$, $u_0 = 79.7~{\rm meV}$, $u_1 = 97.5~{\rm meV}$, $T = 5~\rm K$, and $\bar{\varepsilon}(0) = 4.9$. Data represented by solid red (dashed black) lines have been calculated with the eigenvalues and eigenvectors of the self-consistent Hartree (bare) Hamiltonian Eq.~\eqref{eq:abstract_full_hamiltonian} (Eq.~\eqref{eq:abstract_hamiltonian}). Panel (a) The quantity ${\cal W}^{(0)}$, i.e.~the Drude weight ${\cal D}$ in units of $e^2/\hbar^2$. Panel (b) Ratio between the effective mass $m_{\rm eff}$ defined in Eq.~(\ref{eq:effective_mass}) and the electron mass in vacuum $m_{\rm e}$. (c) Ratio between the effective velocity $v_{\rm eff}$ defined in Eq.~(\ref{eq:effective_velocity}) and the Fermi velocity in monolayer graphene $v_{\rm D}$.}
\end{figure*}
\begin{figure}[t]
\includegraphics[width=\linewidth]{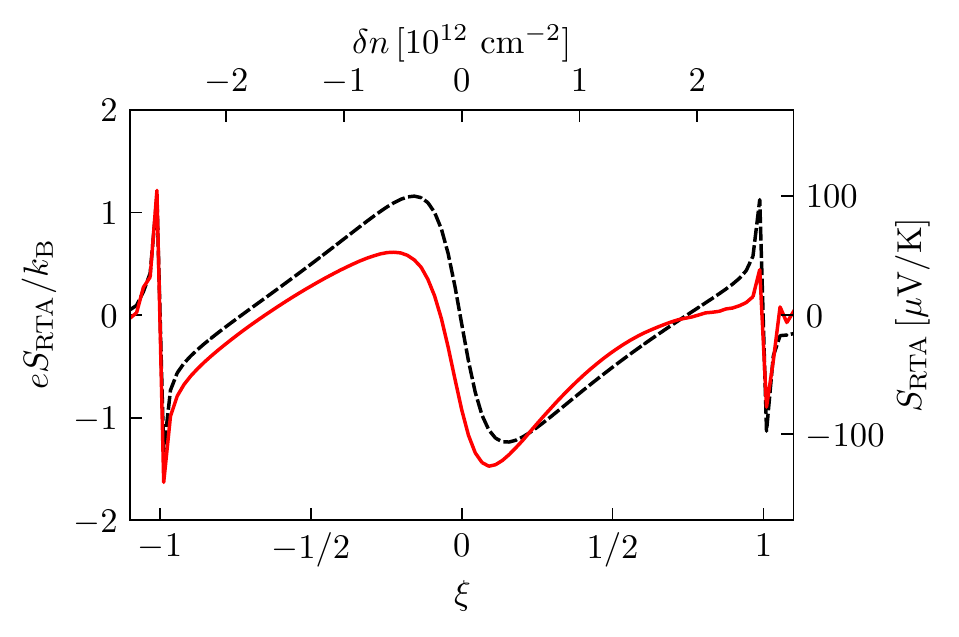}\caption{(Color online)\label{fig:seebeck} Same as in Fig.~\ref{fig:drude_var_filling}---$\theta = 1.05\degree$, $u_0 = 79.7~{\rm meV}$, $u_1 = 97.5~{\rm meV}$, $T = 5~\rm K$, and $\bar{\varepsilon}(0) = 4.9$---but for the Seebeck coefficient $S_{\rm RTA}$ defined in Eq.~(\ref{eq:seebeck_approx}). We performed the same calculations by using the less general Mott formula and found qualitative agreement with the results in this plot.}
\end{figure}
\begin{figure*}[h]
  \includegraphics{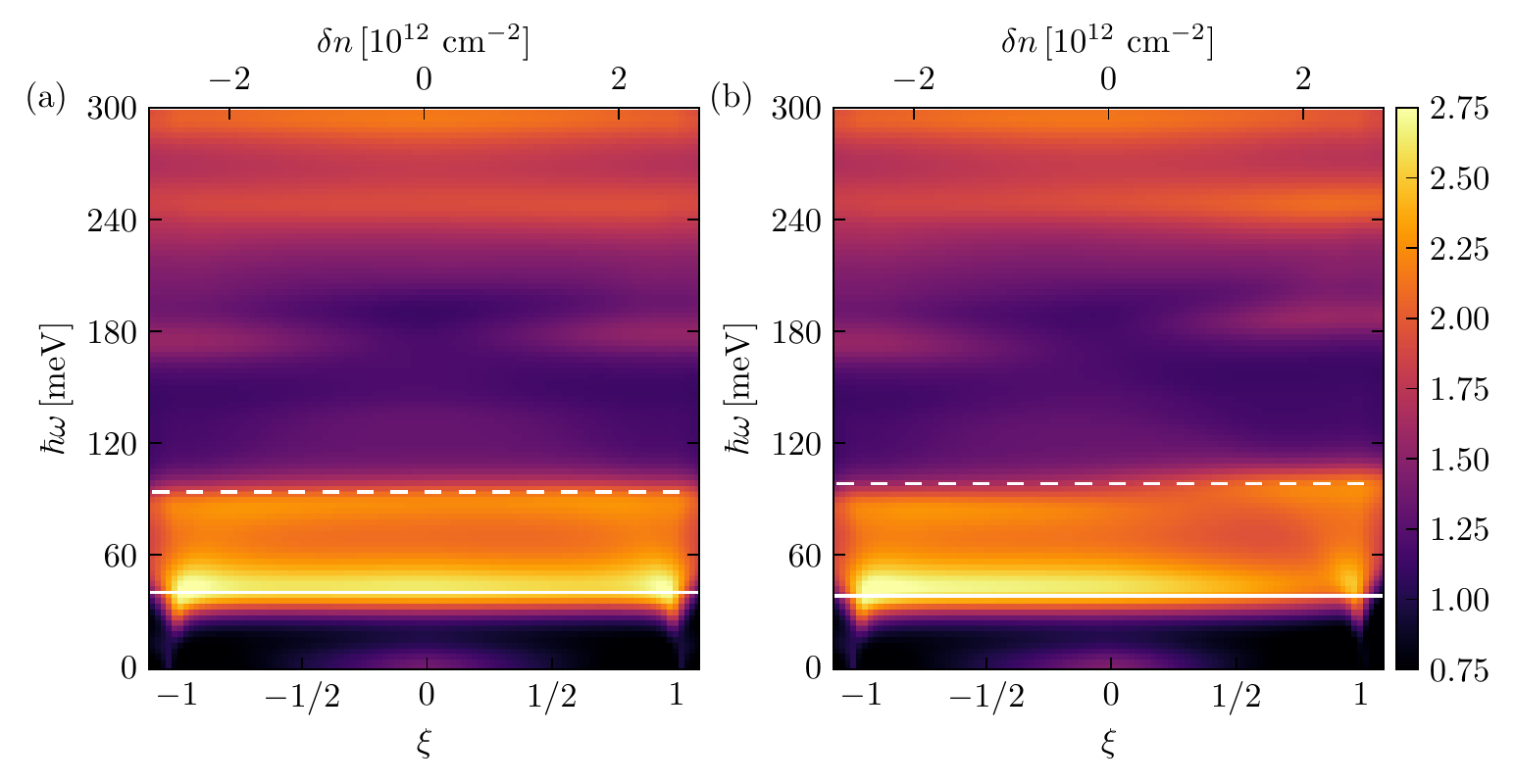}
  \caption{\label{fig:interband_var_filling}(Color online) The inter-band contribution ${\rm Re}[\sigma^{{\rm inter}}(\omega)]$ (in units of $G_0$) to the real part of the optical conductivity, which is  related to optical absorption, is plotted as a function of the photon energy $\hbar\omega$ and carrier density $\delta n$ (or, equivalently, filling factor $\xi$). Data in this plot refer to $\theta = 1.05\degree$, $u_0 = 79.7~{\rm meV}$, $u_1 = 97.5~{\rm meV}$, $T = 5~\rm K$, and $\bar{\varepsilon}(0) = 4.9$. Solid and dashed white lines are placed at energies equal to the gap between the valence flat band and the first non-flat conduction band at the points $\Gamma$ and $K$ in the MBZ, respectively. These energies are associated to the optical transitions marked in panel (a) of Fig.~\ref{fig:bands_var_filling}. Panel (a) ${\rm Re}[\sigma^{\rm inter}(\omega)]$ as calculated from Eq.~\eqref{eq:optical_conductivity_inter} with $\hat{H}({\bm k})$ taken as the non-interacting Hamiltonian (\ref{eq:abstract_hamiltonian}). Panel (b) ${\rm Re}[\sigma^{\rm inter}(\omega)]$ as calculated by taking into account the self-consistent Hartree potential, i.e.~by using Eq.~\eqref{eq:optical_conductivity_inter} with $\hat{H}({\bm k})$ as in Eq.~\eqref{eq:abstract_full_hamiltonian}.}
\end{figure*}
\begin{figure*}[h!]
  \includegraphics{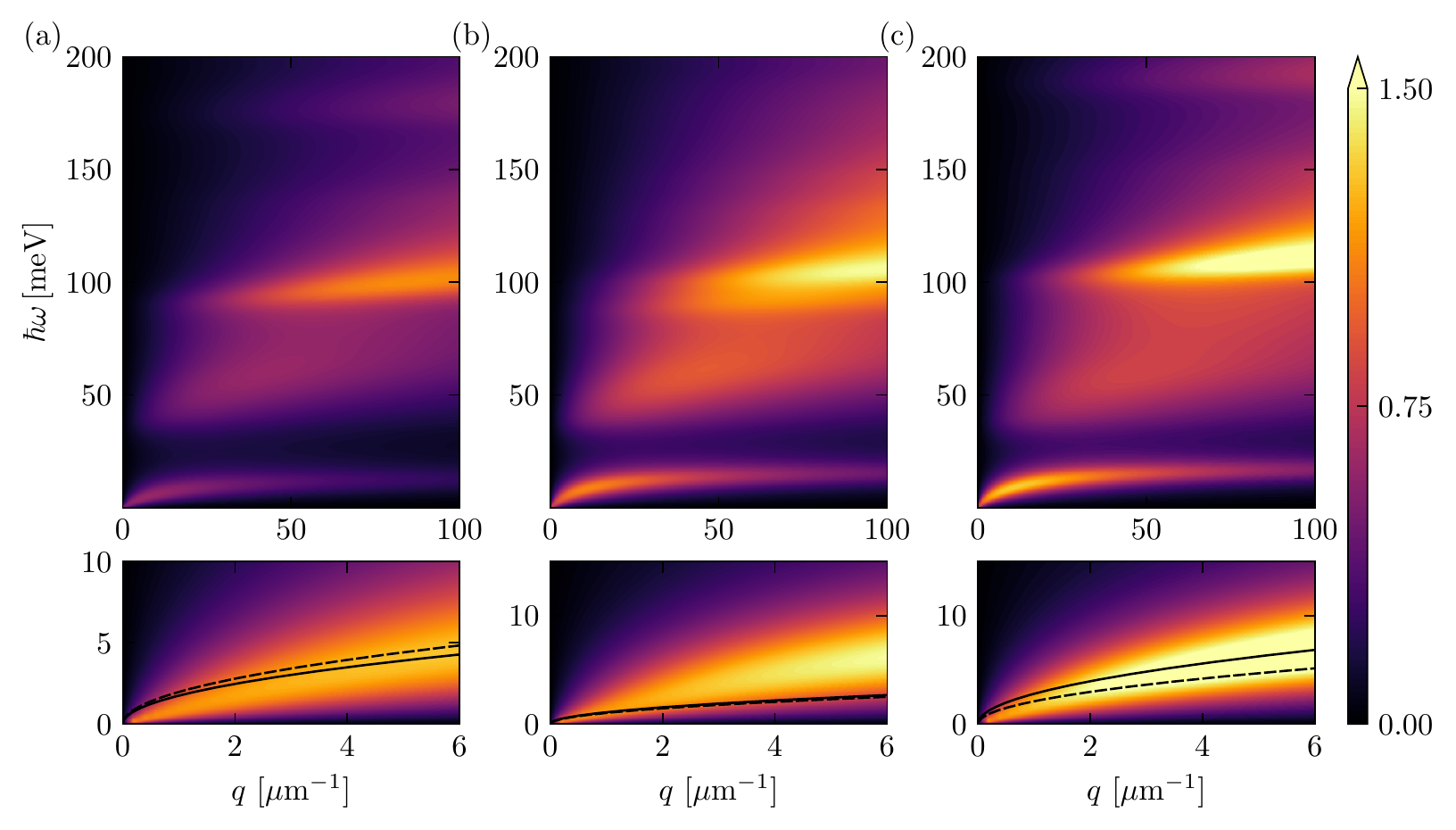}
\caption{\label{fig:loss_function_var_filling} (Color online)  2D plots of the energy loss function $\mathcal{L}(\bm{q},\omega)$, for different values of the filling factor $\xi$ at $\theta = 1.05\degree$, $u_0 = 79.7~{\rm meV}$, $u_1 = 97.5~{\rm meV}$, $T = 5~\rm K$, and $\bar{\varepsilon}(0) = 4.9$: (a) hole doping, $\xi = - 3/4$; (b) CNP, $\xi=0$; (c) electron doping, $\xi = +3/4$. In all the panels, the lower sub panels zoom in on a smaller region of the energy-momentum plane. All 2D plots displayed in this figure have been obtained by using the self-consistent Hartree approximation. The black solid (dashed) lines are the analytical intra-band plasmon dispersions calculated through Eq.~\eqref{eq:sqrt_q_plasmon_dispersion}, making use of eigenvalues and eigenvectors of the self-consistent Hartree (bare) Hamiltonian in Eq.~\eqref{eq:abstract_full_hamiltonian} (Eq.~\eqref{eq:abstract_hamiltonian}), respectively.}
\end{figure*}

The effect of the Hartree potential on the optical conductivity originates from the distortion of both energy bands and wavefunction amplitudes, through the matrix elements of the velocity operator (i.e.~$v_{\alpha, \nu\nu'}  = \hbar^{-1}\langle u_{\bm k \nu}|\partial_{k_\alpha}\hat{H}(\bm k)|u_{\bm k \nu'}\rangle$) in Eqs.~\eqref{eq:drude_weight} and \eqref{eq:optical_conductivity_inter}.
Figure~\ref{fig:drude_var_filling}(a) shows the dependence of $\mathcal{W}^{(0)}$---the Drude weight in units of $e^{2}/\hbar^{2}$---on the filling factor, with (solid red lines) and without (dashed black lines) Hartree self-consistency.
If the chemical potential is within the flat bands, i.e.~if $|\xi| < 1 $, the value of $\mathcal{W}^{(0)}$ is strongly modified by the Hartree potential.
$\mathcal{W}^{(0)}$ is nearly particle-hole symmetric when the Hartree corrections are neglected, but becomes strongly asymmetric when Hartree corrections are accounted for.
In particular, $\mathcal{W}^{(0)}$ reaches its local maxima at filling factors $\xi \simeq \pm 0.7$, with a value of $\sim 7~{\rm meV}$, in the absence of Hartree corrections.
When the Hartree potential is accounted for, the maximum for electron doping (at $\xi \simeq + 0.7$) is $\mathcal{W}^{(0)} \simeq 13~{\rm meV}$, and for hole doping (at $\xi \simeq - 0.7$) is $\mathcal{W}^{(0)} \simeq 5~{\rm meV}$.
Close to the CNP, $\mathcal{W}^{(0)}\simeq 2~{\rm meV}$ both if the Hartree corrections are accounted for or neglected. Note that $\mathcal{W}^{(0)}$ is finite at the CNP because of finite-$T$ effects.
As we shall see later, close to charge neutrality, the Hartree potential only yields modest corrections to the optical properties of TBG.

To give a better physical picture of our results, and to help the comparison with other electronic systems, we can express the Drude weight in the following alternative forms
\begin{equation}\label{eq:effective_mass_and_velocity}
\mathcal{D} \equiv \frac{\pi e^2 |\delta n|}{m_{\rm eff}} \equiv \frac{\pi e^2 v_{\rm eff}^2 N(\mu)}{2}~, 
\end{equation}
where $m_{\rm eff}$ and $v_{\rm eff}$ are an effective mass and velocity, respectively, and $N(\mu)$ is the density of states per unit area, evaluated at the chemical potential. Comparing Eq.~(\ref{eq:effective_mass_and_velocity}) with the formulas in Sect.~\ref{sec:optical_cond} we get
\begin{equation}\label{eq:effective_mass}
m_{\rm eff} = \frac{\pi \hbar^2|\delta n|}{\mathcal{W}^{(0)}}
\end{equation}
and
\begin{equation}\label{eq:effective_velocity}
v_{\rm eff} = \sqrt{\frac{2\mathcal{W}^{(0)}}{\pi \hbar^2 N(\mu)}}~.
\end{equation}
In Fig.~\ref{fig:drude_var_filling} we show plots of these quantities as functions of carrier density.  
As expected, we clearly see that $v_{\rm eff}\ll v_{\rm D}$ in a wide range of carrier densities.

Figure~\ref{fig:seebeck} shows the Seebeck coefficient calculated in the RTA from Eq.~(\ref{eq:seebeck_approx}), as a function of the filling factor. Results obtained from the self-consistent Hartree theory (red curve) are compared with non-interacting results (black dashed curve).
The thermoelectric effect, quantified by the Seebeck coefficient, is one of the main photocurrent generation mechanism in monolayer graphene at room temperature\cite{koppens_nat_nano_2014}, and played an important role in both the study of fundamental phenomena in graphene\cite{lundeberg_nat_mat_2017} and the realization of graphene-based photodetectors\cite{koppens_nat_nano_2014, castilla_nano_lett_2019}. 
Our calculations (see Fig.~\ref{fig:seebeck}) demonstrate that, due to the much slower carrier velocity, TBG maintains a significant thermoelectric effect even at cryogenic temperatures $\approx 5\rm K$, making low-temperature photocurrent spectroscopy a useful technique to study TBG close to the transition to correlated states.

In Fig.~\ref{fig:interband_var_filling} we display the real part of the inter-band optical conductivity as calculated from Eq.~\eqref{eq:optical_conductivity_inter}. The imaginary part can be straightforwardly obtained from the Kramers-Kronig relation~\cite{Giuliani_and_Vignale}. 

The quantity ${\rm Re}[\sigma^{{\rm inter}}(\omega)]$ is related to the inter-band optical absorption at an incident photon energy $\hbar\omega$. 
If the matrix elements $v_{\alpha,\nu\nu'}$ are non-zero for symmetry reasons, peaks are expected in ${\rm Re}[\sigma^{{\rm inter}}(\omega)]$ when the photon energy matches a vertical inter-band transition, i.e.~when $\epsilon_{\bm{k}\nu} - \epsilon_{\bm{k}\nu^{\prime}} + \hbar\omega \approx 0$ in Eq.~\eqref{eq:optical_conductivity_inter}. 
Multiple distinct peaks of ${\rm Re}[\sigma^{{\rm inter}}(\omega)]$ are visible in Fig.~\ref{fig:interband_var_filling}, two of which are highlighted explicitly. 
At the very bottom of the two panels, for $\hbar\omega \lesssim 10~{\rm meV}$, the lighter spot close to the CNP stems from a weak inter-flat-band contribution to the optical conductivity. 
Increasing $\omega$, ${\rm Re}[\sigma^{{\rm inter}}(\omega)]$ decreases until $\hbar \omega \simeq 40~{\rm meV}$, where it reaches its absolute maximum. 
The position of this peak is pretty much identical and filling-independent in both panels, whereas its intensity is slightly different in the two panels, with a filling-dependent intensity for the case of the results obtained with the Hartree self-consistency, panel b). 
The optical transitions associated with this peak are due to electrons with momenta close to the $\Gamma$ point in the MBZ that are excited by photons from the valence flat band to the first non-flat conduction band. 
This optical transition is highlighted with a solid arrow in panel (a) of Fig.~\ref{fig:bands_var_filling}. 
Part of the spectral weight of this peak is also due to transitions from the first non-flat valence band to the conduction flat band. 
The second notable peak in ${\rm Re}[\sigma^{\rm inter}(\omega)]$ occurs at $\hbar\omega \simeq 95~{\rm meV}$ and is associated to optical transitions between the same bands involved in the previously discussed peak, albeit for electrons in the vicinity of the corners of the MBZ, as showed by the dashed arrow in panel (a) of Fig.~\ref{fig:bands_var_filling}. 

The effect of the Hartree self-consistency on the inter-band contribution ${\rm Re}[\sigma^{\rm inter}(\omega)]$ to the optical conductivity is mostly appreciable in the vicinity of its peaks. 
The intensity of the strongest peak becomes filling-dependent when the Hartree corrections are taken into account, with higher intensity at negative values of $\xi$, i.e.~for hole doping. The second most-intense peak, which, as stated above, originates from transitions occurring near the $K$ point in the MBZ, is not affected in its intensity by the Hartree corrections. Nonetheless, switching from negative to positive filling factors, the energy at which the peak occurs varies slightly. This can be understood by recalling that, as discussed above, the non-flat bands are rigidly shifted by the Hartree potential, whereas the flat bands are unaffected by $V_{\rm H}$ in the vicinity of the $K$ point in the MBZ.

In Fig.~\ref{fig:loss_function_var_filling} we illustrate the dependence of the loss function on the filling factor, for the same values of $\xi$ as in Fig.~\ref{fig:bands_var_filling} and for the same parameters $u_0$, $\theta$, and $T$. 
$\mathcal{L}(\bm{q},\omega)$ encodes both inter- and intra-band contributions, as already discussed for the conductivity $\sigma(\omega)$. 
The color plots in Fig.~\ref{fig:loss_function_var_filling} have been obtained by employing the fully self-consistent Hartree model, Eq.~\eqref{eq:abstract_full_hamiltonian}. 
For each of the columns in Fig.~\ref{fig:loss_function_var_filling}, the upper panel displays $\mathcal{L}(\bm{q},\omega)$ in a range of energies and wave vectors where inter-band plasmons are excited~\cite{stauber_nanolett_2016}. 
Conversely, the lower panels are a zoom at small $\omega$ and $q$. In the latter, ordinary intra-band plasmons~\cite{Giuliani_and_Vignale} are clearly visible, whose dispersion relation admits a simple analytical description. 
The plasmon peaks, indeed, stem from zeroes of the longitudinal dielectric function, Eq.~\eqref{eq:non_local_epsilon}. 
Plasmon dispersions originating from intra-band processes are easily extracted by plugging the value of the intra-band optical conductivity \eqref{eq:optical_conductivity_intra} into Eq.~\eqref{eq:non_local_epsilon}. After straightforward manipulations, we  reach the usual~\cite{Giuliani_and_Vignale} 2D intra-band plasmon dispersion relation
\begin{equation}\label{eq:sqrt_q_plasmon_dispersion}
  \hbar\omega_{\rm pl}(q\to 0) = \sqrt{\frac{2e^{2} {\cal W}^{(0)}q}{\bar{\varepsilon}(0)}}~.
\end{equation}

In the lower panels of Fig.~\ref{fig:loss_function_var_filling} we have also plotted the previous equation using the values of $\mathcal{W}^{(0)}$ computed both with and without Hartree corrections. 
Away from the CNP---panels (a) and (c) of Fig.~\ref{fig:loss_function_var_filling}---the loss function has a clearly distinguishable peak dispersing as predicted by Eq.~(\ref{eq:sqrt_q_plasmon_dispersion}). The two analytical dispersion relations are different because they depend on the value of $\mathcal{W}^{(0)}$, which, as we have seen before, is modified by the Hartree potential with respect to the bare value when TBG is doped away from the CNP. Recalling that the color plots refer to the fully self-consistent Hartree theory, it is no surprise to see that the intra-band plasmon mode observed as a peak in $\mathcal{L}(\bm{q},\omega)$ at small $q$ and $\omega$ is centered around the dispersion relation calculated with the fully self-consistent Hartree value of $\mathcal{W}^{(0)}$, i.e.~around the solid black line.

A completely different behavior is observed at the CNP. In this case the loss function displays a well defined plasmon branch which, however, does not follow the approximate analytic plasmon dispersion in Eq.~(\ref{eq:sqrt_q_plasmon_dispersion}). This is readily explained by remembering that the analytic plasmon dispersion presented above describes collective excitations arising from intra-band processes. At the CNP, the Fermi surface shrinks down to a single point, and intra-band collective modes can originate only from finite-temperature effects (i.e.~thermally-excited quasiparticles). Albeit the present calculations are carried out at a finite temperature, $T= 5~{\rm K}$, the intra-flat-band plasmon branch due to thermally excited quasiparticles is not a clearly distinguishable component of the low-energy loss function. Rather, the low-energy plasmon branch visible at the CNP stems from optical transitions between the flat bands. This follows from simple energetic considerations. Since the characteristic energy scale of this plasmon is $\lesssim 20~{\rm meV}$, the inter-band processes from which it originates are bound to occur in the manifold of nearly-flat bands. This is justified by observing that exciting electrons onto the higher energy bands would require an energy $\hbar\omega > 20~{\rm meV}$. We note that at $\xi=0$ the two analytical dispersion relations shown at the bottom of panel (b) are almost identical. This is because $\mathcal{W}^{(0)}$---as previously mentioned---is unaffected by the Hartree potential at the CNP.

For any of the values of the filling factor, there is also another quite noticeable peak in $\mathcal{L}(\bm{q},\omega)$ at energies $\hbar \omega \approx 100~{\rm meV}$. This an inter-band plasmon, analogous to the one measured in Ref.~\onlinecite{hesp_arxiv_2019} at $\theta=1.35\degree$. It starts off at a finite wave vector $\approx 5~{\rm \mu m}^{-1}$ and its position in the $\omega$-$q$ plane is just weakly affected  by the filling factor $\xi$.

The optical transitions responsible for this inter-band plasmon are the ones occurring at the energy highlighted by the dashed white line in panel (b) of Fig.~\ref{fig:interband_var_filling}. At $\theta=1.05\degree$ and for the values of the parameter $u_0$ chosen in this Section, this inter-band plasmon originates from processes occurring near the corners of the MBZ. 

We finally wish to stress that electron-hole attraction effects (i.e.~excitonic effects), which are missed by the RPA theory we are employing in this work, may alter our results on inter-band plasmons, even at relatively small values of $q$. Much more work is needed to quantify such excitonic effects in TBG, the minimal theory that captures these effects being the TDHF approximation, briefly mentioned in Sect.~\ref{sect:intro}.

\begin{figure*}[t]
  \includegraphics[width =\linewidth]{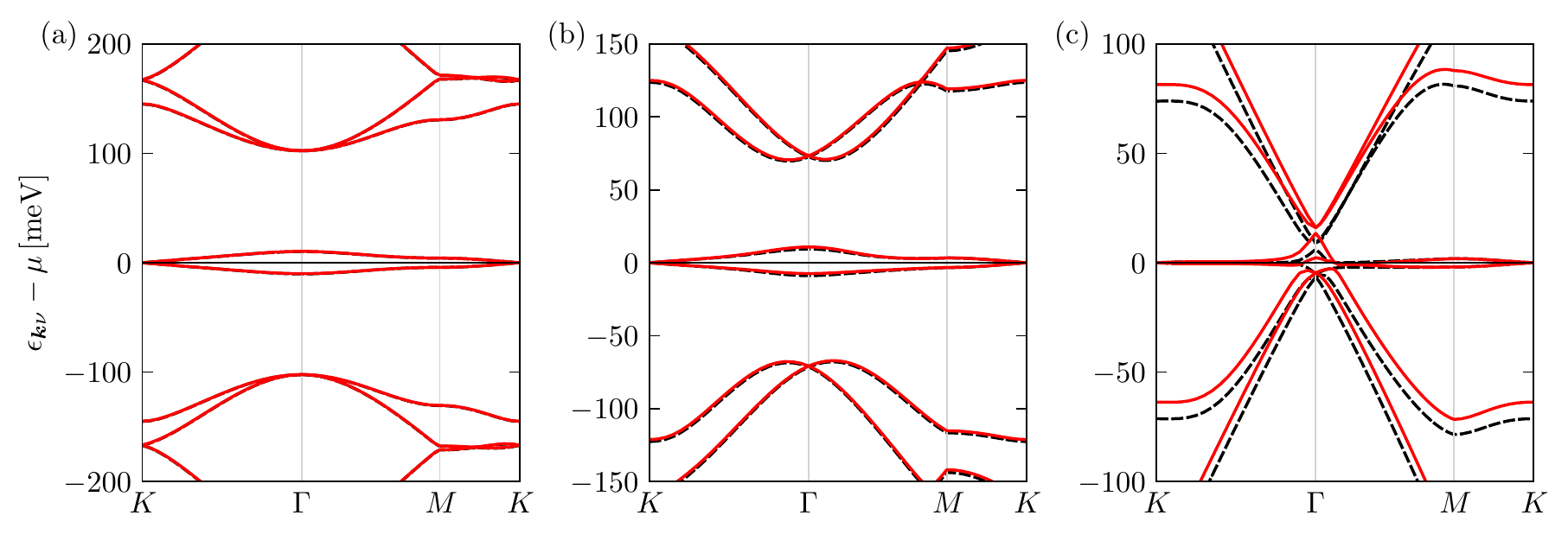}
  \caption{\label{fig:bands_var_delta} (Color online) TBG energy bands for different values of $u_0$ are plotted along the $K$-$\Gamma$-$M$-$K$ path in the MBZ. Data in this figure refer to $\theta = 1.05\degree$, $u_1 = 97.5~{\rm meV}$, $T = 5~{\rm K}$, $\xi=0$, and $\bar{\varepsilon}(0) = 4.9$. Color coding and line styles have the same meaning as in Fig.~\ref{fig:bands_var_filling}. Different panels refer to different values of the intra-sublattice inter-layer tunneling energy $u_0$. Panel (a) $u_0 = 0~{\rm meV}$ (as in Ref.~\onlinecite{tomarken_prl_2019}). Panel (b) $u_0 = 48.2~{\rm meV}$. Panel (c) $u_0 = u_1 = 97.5~{\rm meV}$ (as in  Ref.~\onlinecite{bistritzer_pnas_2011}).}
\end{figure*}
\subsection{Dependence on the intra-sublattice inter-layer tunneling energy $u_0$}\label{subsec:var_delta}

In this Section we present numerical results for $\sigma(\omega)$ and $\mathcal{L}(\bm{q},\omega)$ obtained by changing the intra-sublattice inter-layer hopping energy $u_0$. As in Sect.~\ref{subsec:var_filling}, the inter-sublattice inter-layer hopping energy has been fixed at $u_1 = 97.5~{\rm meV}$, the twist angle at $\theta = 1.05\degree$, and the temperature at $T=5~{\rm K}$. We here study the dependence on $u_0$ only at the CNP, i.e.~at $\xi=0$.

It is known~\cite{tomarken_prl_2019} that for, $u_0 = 0$, the flat bands' bandwidth at the magic angle is exactly zero throughout the whole MBZ. Since $\theta = 1.05\degree$ is close to but not exactly the magic angle, the flat bands' bandwidth is non-zero even at $u_0=0$. It is also known~\cite{tomarken_prl_2019} that TBG at small values of $u_0$ hosts large (i.e.~on the order of $\approx 100~{\rm meV}$) energy gaps between the flat bands and the ``remote" conduction/valence bands. These gaps therefore provide a rough estimate of the energy scales at which optical transitions occur. This is going to be quite evident both in the optical conductivity and loss function calculated at $u_0=0$.
\begin{figure*}[t]
  \includegraphics{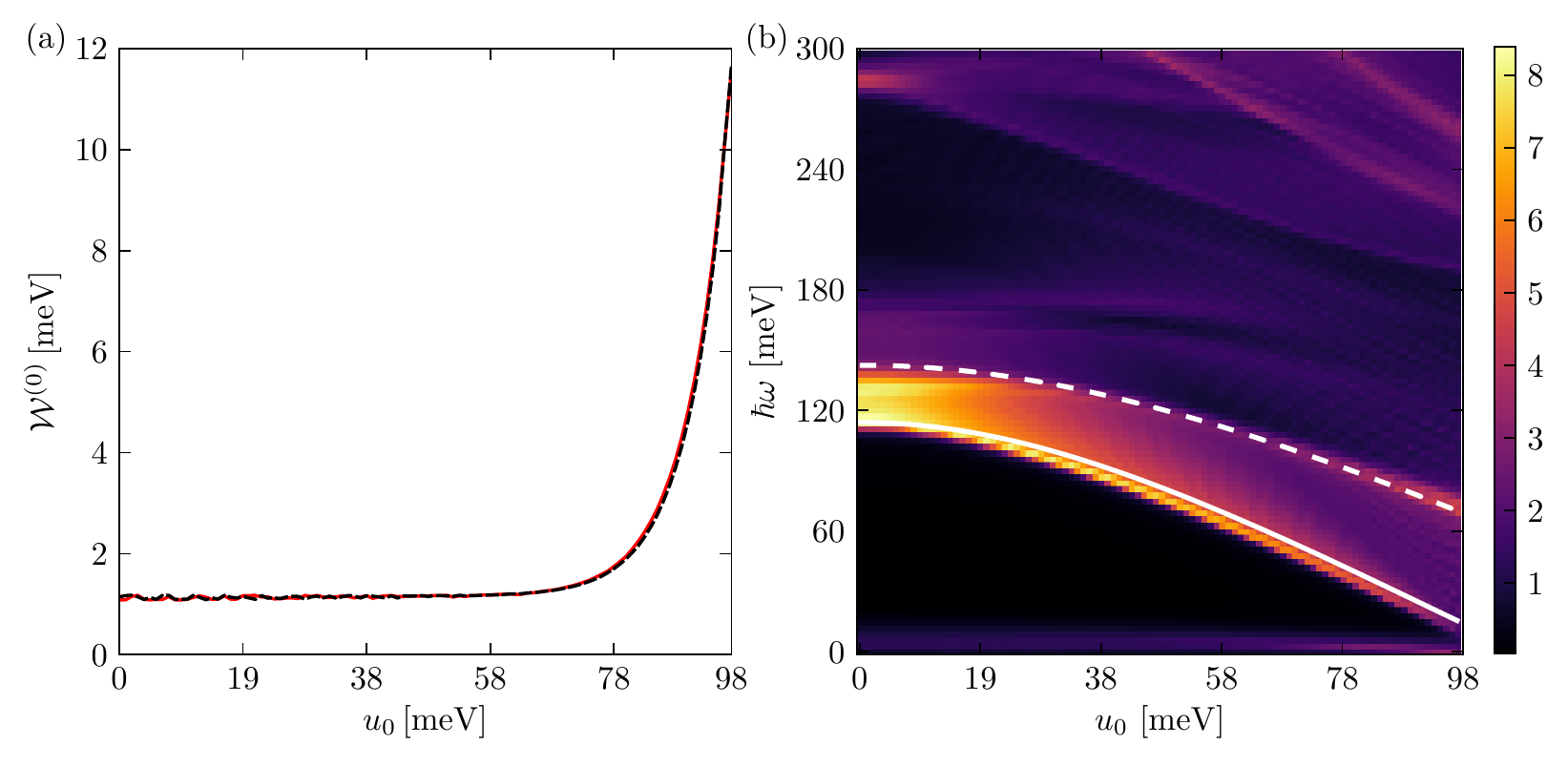}
  \caption{\label{fig:optical_conductivity_var_delta} (Color online) Panel (a) Drude weight in units of $e^{2}/\hbar^{2}$ as a function of $u_0$, for $\theta= 1.05\degree$, $u_1 = 97.5~{\rm meV}$, $T = 5~{\rm K}$, $\xi=0$, and $\bar{\varepsilon}(0) = 4.9$. Color coding and line styles have the same meaning as in panel (a) of Fig.~\ref{fig:drude_var_filling}. Panel (b) The quantity ${\rm Re}[\sigma^{\rm inter}(\omega)]$ (in units of $G_0$), related to the optical absorption, is plotted as a function of $\hbar\omega$ and $u_0$. Results in this panel have been obtained by employing the Hartree self-consistent approximation and refer to $\theta = 1.05\degree$, $u_1 = 97.5~{\rm meV}$,  $T = 5~{\rm K}$, and $\xi=0$. Solid and dashed white lines are placed at energies equal to the gap between the valence flat band and the first non-flat conduction band at the point $\Gamma$ and $K$ in the MBZ, respectively. These lines are associated to the optical transitions marked in Fig.~\ref{fig:bands_var_filling}(a).}
\end{figure*}

In Fig.~\ref{fig:bands_var_delta} we display the energy bands of TBG at different values of $u_0$. At $u_0 = 0~{\rm meV}$, the Hartree corrections on the band structure are negligible and the energy gap between the flat bands and the adjacent bands is on the order of $\approx 100~{\rm meV}$. At $u_0 = 48.2~{\rm meV}$, again, the Hartree potential leaves the bare energy bands almost unchanged. In this case, however, the energy gap between the flat bands and the adjacent bands is $\approx 120~{\rm meV}$ near the $K$ point and $\approx 75~{\rm meV}$ near the $\Gamma$ point.  Finally, at $u_0 = u = 97.5~{\rm meV}$, the Hartree potential manifests as an upward bending of the flat bands, most noticeably near the $\Gamma$ point, whereas the energy gap between flat bands and adjacent  bands is $\lesssim 5~{\rm meV}$ at the $\Gamma$ point and $\approx 75~{\rm meV}$ at the $K$ point. We remind the reader that TBG bands at $u_0 = 79.7~{\rm meV}$, which is the value predicted for corrugated TBG~\cite{koshino_prx_2018,lucignano_prb_2019}, and $\xi=0$ can be found in panel (b) of Fig.~\ref{fig:bands_var_filling}. 

An important remark is now in order. 
Even though the Hartree contribution distorts the bare bands, the energy gaps between flat bands and adjacent conduction/valence bands are virtually the same as in  the case of the bare bands. This is another manifestation of the previously noted fact that, close to zero filling, the optical properties are qualitatively unaffected by the Hartree potential. Conversely, the value of $u_0$ dramatically alters the energies at which optical transitions with large spectral weight occur.

In Fig.~\ref{fig:optical_conductivity_var_delta} we show ${\cal W}^{(0)}$, i.e.~the Drude weight in units of $e^2/\hbar^{2}$, and the real part ${\rm Re}[\sigma^{\rm inter}(\omega)]$ of the inter-band optical conductivity. We note that $\mathcal{W}^{(0)}$ is an increasing function of $u_0$. This follows from the fact that $\mathcal{W}^{(0)}$, whose microscopic expression can be obtained from Eq.~\eqref{eq:drude_weight} by setting $\alpha = \beta$, depends on the derivative of the bands with respect to ${\bm k}$, i.e.~on $|\langle \bm{k}\nu |\partial_{k_{\alpha}}\hat{H}(\bm{k})|\bm{k}\nu\rangle|^{2} = |\partial_{k_{\alpha}} \epsilon_{\bm{k}\nu}|^{2}$. Now, as shown in Fig.~\ref{fig:bands_var_delta}, the flat bands at $u_0 = 0$ vary more smoothly throughout the MBZ with respect to the bands evaluated at finite $u_0$. In the latter case, we note a sudden variation of the band dispersion in the vicinity of the $\Gamma$ point.  
Once again, since we are at the CNP, the quantity ${\cal W}^{(0)}$ calculated in the fully self-consistent Hartree approximation is practically indistinguishable with respect to the bare result, as evident from panel (a) of Fig.~\ref{fig:optical_conductivity_var_delta}. In panel (b) of Fig.~\ref{fig:optical_conductivity_var_delta} we therefore plot the real part ${\rm Re}[\sigma^{\rm inter}(\omega)]$ of the inter-band contribution to the optical conductivity calculated in the Hartree approximation. We clearly see that ${\rm Re}[\sigma^{\rm inter}(\omega)]$ shows a very interesting dependence on $u_0$, with its peaks shifting sensibly with it. As in Fig.~\ref{fig:interband_var_filling}, the solid white line is the energy separation between the valence flat band and the first non-flat conduction band at the $\Gamma$ point, whereas the dashed white line is the energy separation between the same pairs of bands, albeit evaluated at the $K$ point in the MBZ. The position of the peak corresponding to the optical transition at the $\Gamma$ point decreases monotonically with $u_0$ from a maximum of $\hbar \omega \simeq 120~{\rm meV}$ at $u_0 = 0~{\rm meV}$ to a minimum of $\hbar \omega \simeq 5~{\rm meV}$ at $u_0 = u= 97.5~{\rm meV}$. These energy values are recovered also by looking at the band structures in Fig.~\ref{fig:bands_var_delta}. A similar, monotonically decreasing behavior is followed by the peaks associated to the optical transitions near the $K$ point in the MBZ. In this case, the position of the peak is $\hbar \omega \simeq 150~{\rm meV}$ at $u_0 = 0~{\rm meV}$ and $\hbar \omega \simeq 70~{\rm meV}$ at $u_0 = 97.5~{\rm meV}$. As a final note on the inter-band optical conductivity, we stress that the intra-sublattice hopping energy scale $u_0$ is responsible also for sensible shifts in the position of optical transitions at higher energies, up to hundreds of meV (see Fig.~\ref{fig:optical_conductivity_var_delta}).

\begin{figure*}[h!]
  \includegraphics{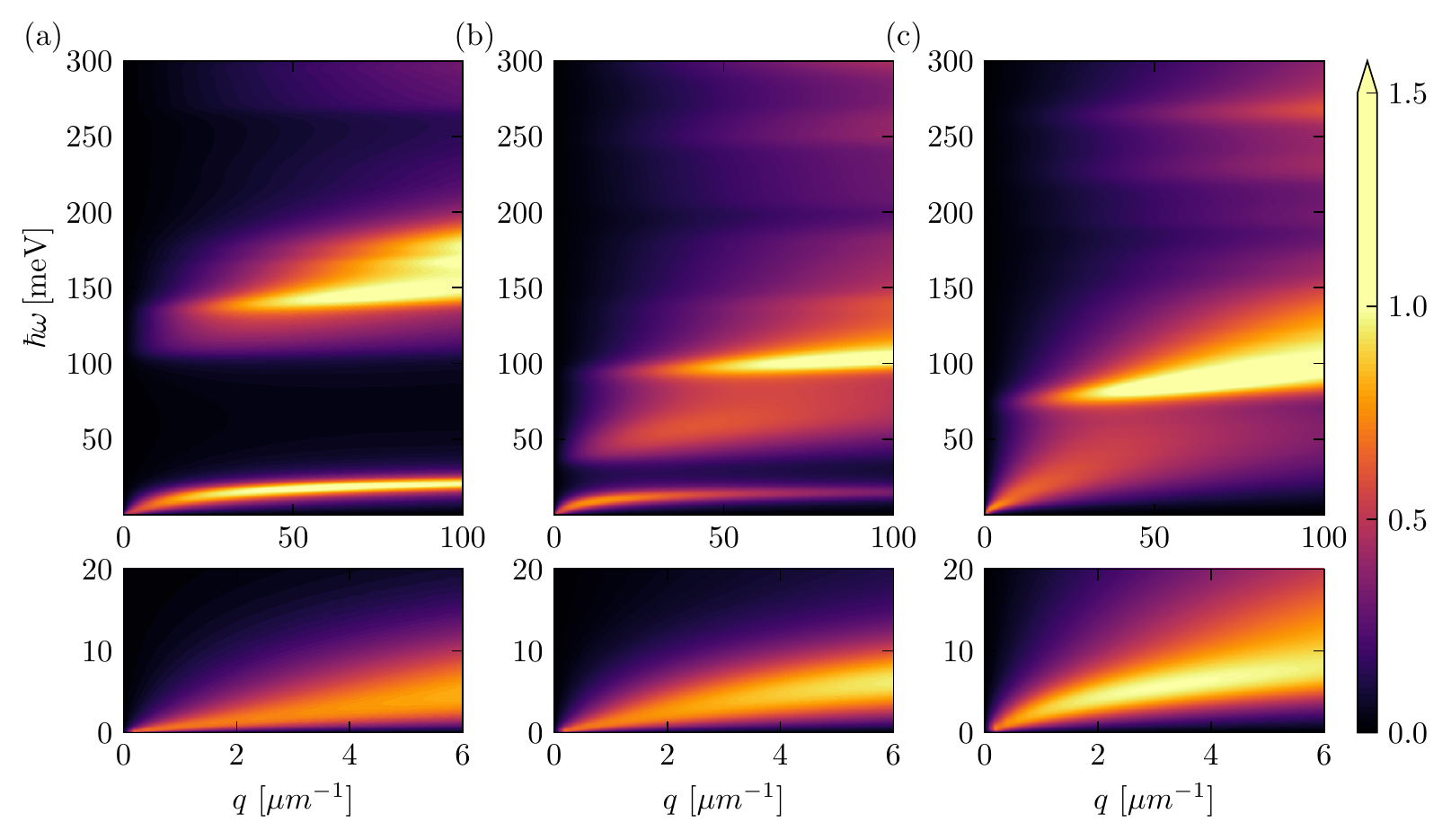}
  \caption{\label{fig:loss_var_delta} (Color online)  2D plots of the loss function $\mathcal{L}(\bm{q},\omega)$ for different values of the intra-sublattice hopping energy $u_0$ at $\theta = 1.05\degree$, $u_1 = 97.5~{\rm meV}$, $T = 5~{\rm K}$, $\xi = 0$, and $\bar{\varepsilon}(0) = 4.9$: (a) $u_0 = 0$ as in Ref.~\onlinecite{tomarken_prl_2019}; (b) $u_0 = 48.2~{\rm meV}$; (c) $u_0 = u_1 = 97.5~{\rm meV}$ as in Ref.~\onlinecite{bistritzer_pnas_2011}. In all the panels, the lower sub panels zoom in on a smaller region of the energy-momentum plane. Data displayed in this figure have been obtained by employing the self-consistent Hartree approximation at the CNP ($\xi=0$).}
\end{figure*}

The loss function, evaluated for different values of $u_0$, is showed in Fig.~\ref{fig:loss_var_delta}. As before, the upper panels display $\mathcal{L}(\bm{q},\omega)$ at energy scales which are suited to inspect collective excitations originating from inter-band processes. Inter-band plasmon branches are clearly visible at energies $\hbar\omega >  50~{\rm meV}$ and are analogous to the ones experimentally measured in Ref.~\onlinecite{hesp_arxiv_2019}. The position of these branches in the $\omega$-$q$ plane is fairly sensitive to the value of the parameter $u_0$. An inter-band plasmon with characteristic excitation energy $\hbar\omega \approx 140~{\rm meV}$ at $u_0 =0~{\rm meV}$ drops down in energy to $\hbar\omega \approx 80~{\rm meV}$ at $u_0 = 97.5~{\rm meV}$. As before, at excitation energies $\lesssim 20~{\rm meV}$ there is a quite evident plasmon branch, which originates from inter-flat-band optical transitions. Starting from $u_0=0$, the corresponding peak in the loss function is well defined up to $u_0 = 48.2~{\rm meV}$, whereas it rapidly broadens in the limit $u_0 = u_1 = 97.5~{\rm meV}$, as shown in the corresponding upper panel. The lower panels of Fig.~\ref{fig:loss_var_delta} illustrate  the loss function at small $q$ and $\omega$. The analytical plasmon dispersion Eq.~\eqref{eq:sqrt_q_plasmon_dispersion} is not shown because, as discussed in the previous Section, it is suited to describe intra-band plasmons. At charge neutrality, such intra-band excitations can arise only from finite-temperature effects, and in the present case ($T=5~{\rm K}$) it is pratically impossible to clearly identify their contribution to the loss function.
\subsection{Dependence on the twist angle}\label{subsec:var_theta}
\begin{figure*}[h!]
  \includegraphics{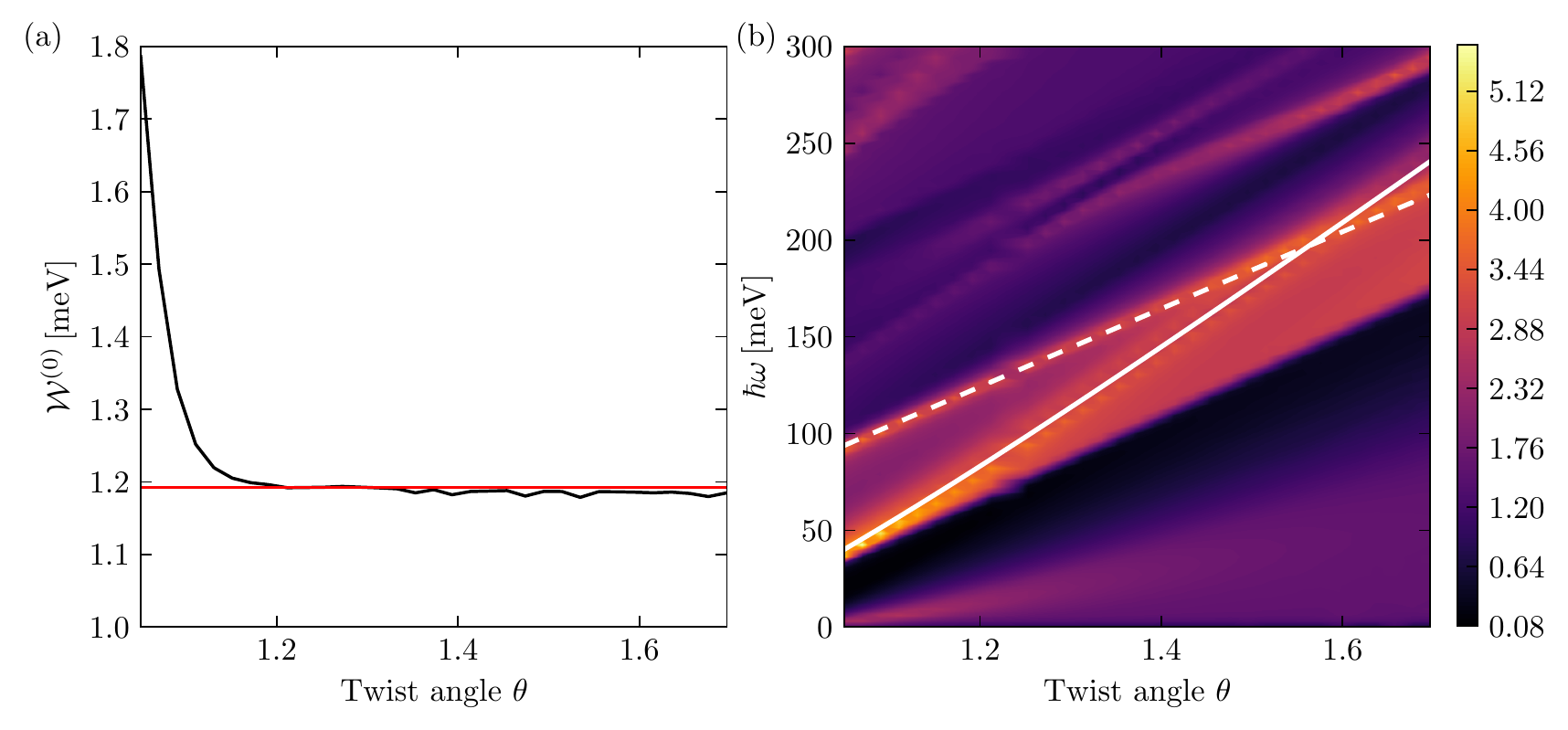}
  \caption{\label{fig:optical_conductivity_var_theta} (Color online) (a) The Drude weight (in units of $e^{2}/\hbar^{2}$) is plotted as a function of the twist angle $\theta$. The solid black trace denotes data calculated via Eq.~\eqref{eq:drude_weight} at $\xi=0$, $u_0=79.7~{\rm meV}$, $u_1 = 97.5~{\rm meV}$, $T = 5~\rm K$, and $\bar{\varepsilon}(0) = 4.9$, and obtained by making use of the eigenvalues and eigenvectors of self-consistent Hartree Hamiltonian~\eqref{eq:abstract_full_hamiltonian}. The solid red line is the value of $\mathcal{W}^{(0)}$ calculated analytically for a linear energy dispersion relation. (b) The quantity ${\rm Re}[\sigma^{\rm inter}(\omega)]$ (in units of $G_0$) is plotted as a function of $\hbar\omega$ and $\theta$. Data in this plot have been obtained by setting $\xi=0$, $u_0=79.7~{\rm meV}$, $u_1 = 97.5~{\rm meV}$, and $T = 5~\rm K$, and calculated from Eq.~\eqref{eq:optical_conductivity_inter} with the fully self-consistent Hamiltonian \eqref{eq:abstract_full_hamiltonian}. Solid and dashed white lines are placed at energies equal to the gap between the valence flat band and the first non-flat conduction band at the point $\Gamma$ and $K$ in the moir\'e Brillouin zone, respectively. These lines are associated to the optical transitions marked in Fig.~\ref{fig:bands_var_filling}(a).}
\end{figure*}

We now move on to discuss the optical conductivity and loss function of TBG as functions of the twist angle $\theta$. It is known~\cite{lopes_prb_2012,bistritzer_pnas_2011} that, for a fraction of their bandwidth, TBG's low-energy bands disperse linearly, akin to the ones of monolayer graphene, albeit with a renormalized Fermi velocity. The linear energy dispersion of TBG, however, extends over a fraction of the low-energy bands' bandwidths that decreases very rapidly as a function of $\theta$.  

The dependence of $\sigma(\omega)$ on $\theta$, down to $\theta \gtrsim 2.0 \degree$, has been studied in  Ref.~\onlinecite{moon_prb_2013}. For this reason, we will focus on $\theta \lesssim 2.0\degree$. We set $\xi=0$, $u_1 = 97.5~{\rm meV}$, and $u_0 = 79.7~{\rm meV}$. As in the previous Sections, $T= 5~{\rm K}$. 

The dependence of the band structure of TBG on $\theta$ has been extensively discussed in the literature~\cite{lopes_prb_2012, bistritzer_pnas_2011,carr_prresearch_2019}. The bandwidth of the ``flat bands", i.e.~the bands closer to the CNP at $1.05\degree$, increases very rapidly with $\theta$, becoming~\cite{moon_prb_2013} $\approx 500~{\rm meV}$ at $\theta \approx 2.5\degree$, i.e.~the two ``flat" bands extend over a total energy range of $\approx 1~{\rm eV}$. In light of this, from now on we will refer to these bands as first conduction and valence bands.

In Fig.~\ref{fig:optical_conductivity_var_theta} we show the Drude weight in units of $e^{2}/\hbar^{2}$, i.e.~$\mathcal{W}^{(0)}$, and the real part ${\rm Re}[\sigma^{{\rm inter}}(\omega)]$ of the inter-band optical conductivity as functions of $\theta$. The quantity $\mathcal{W}^{(0)}$ is a monotonically decreasing function of the twist angle, approaching an asymptotic value at large $\theta$, which can be calculated analytically. A straightforward calculation, indeed, shows that, at the CNP, the value of $\mathcal{W}^{(0)}$ for linear energy bands is $\mathcal{W}^{(0)} = gk_{\rm B}T\log (2)$, independent of the Fermi velocity. The value of $k_{\rm B}T \approx 0.4~{\rm meV}$, chosen in our numerical calculations, is much smaller than the bandwidth of the valence and conduction bands, especially so for the case of $\theta \gtrsim 1.2\degree$ (see Fig.~\ref{fig:optical_conductivity_var_theta}). The quantity $\mathcal{W}^{(0)}$, thus, converges to the asymptotic limit $gk_{\rm B}T\log (2)$ when the value of $k_{\rm B}T$ is much smaller than the energy scale over which the bands are linear. On the other hand, at smaller twist angles---and generally speaking when $k_{\rm B}T$ is larger than or comparable to the energy range over which the first valence and conduction bands are linear---$\mathcal{W}^{(0)}$ increases. In panel (a) of Fig.~\ref{fig:optical_conductivity_var_theta}, it is evident that already at $T = 5~{\rm K}$, the Drude weight of TBG (in units of $e^{2}/\hbar^{2}$) has values that are quite different from the ones expected for a material with linearly-dispersing energy bands. This effect is expected to be enhanced by temperature, i.e.~for higher $T$, the value of $\mathcal{W}^{(0)}$ is expected to converge to $gk_{\rm B}T\log (2)$ at larger twist angles.

As we discussed earlier, ${\rm Re}[\sigma^{{\rm inter}}(\omega)]$ shows peaks at energies $\hbar\omega$ at which the denominator in Eq.~\eqref{eq:optical_conductivity_inter} is minimal, i.e.~when $\hbar\omega + \epsilon_{\bm{k}\nu} - \epsilon_{\bm{k}\nu^{\prime}} \approx 0$. The energies at which those peaks occur increase monotonically with the twist angle. In panel (b) of Fig.~\ref{fig:optical_conductivity_var_theta} we have marked with solid and dashed white lines the excitation energies of the optical transitions occurring near the $\Gamma$ and $K$ points of the MBZ, respectively. Around $\theta \approx 1.55\degree$ these lines cross, meaning that the energy distance between the valence band and the second conduction band is wider at $\Gamma$ than at $K$.  As in the case of variable intra-sub-lattice hopping energy, the positions of the peaks of the optical conductivity change with $\theta$, in a wide range of energies. Fig.~\ref{fig:optical_conductivity_var_theta}(b) shows that these modifications occur up to energies $ \hbar \omega \simeq 300~{\rm meV}$. 
\begin{figure*}[h!]
  \includegraphics{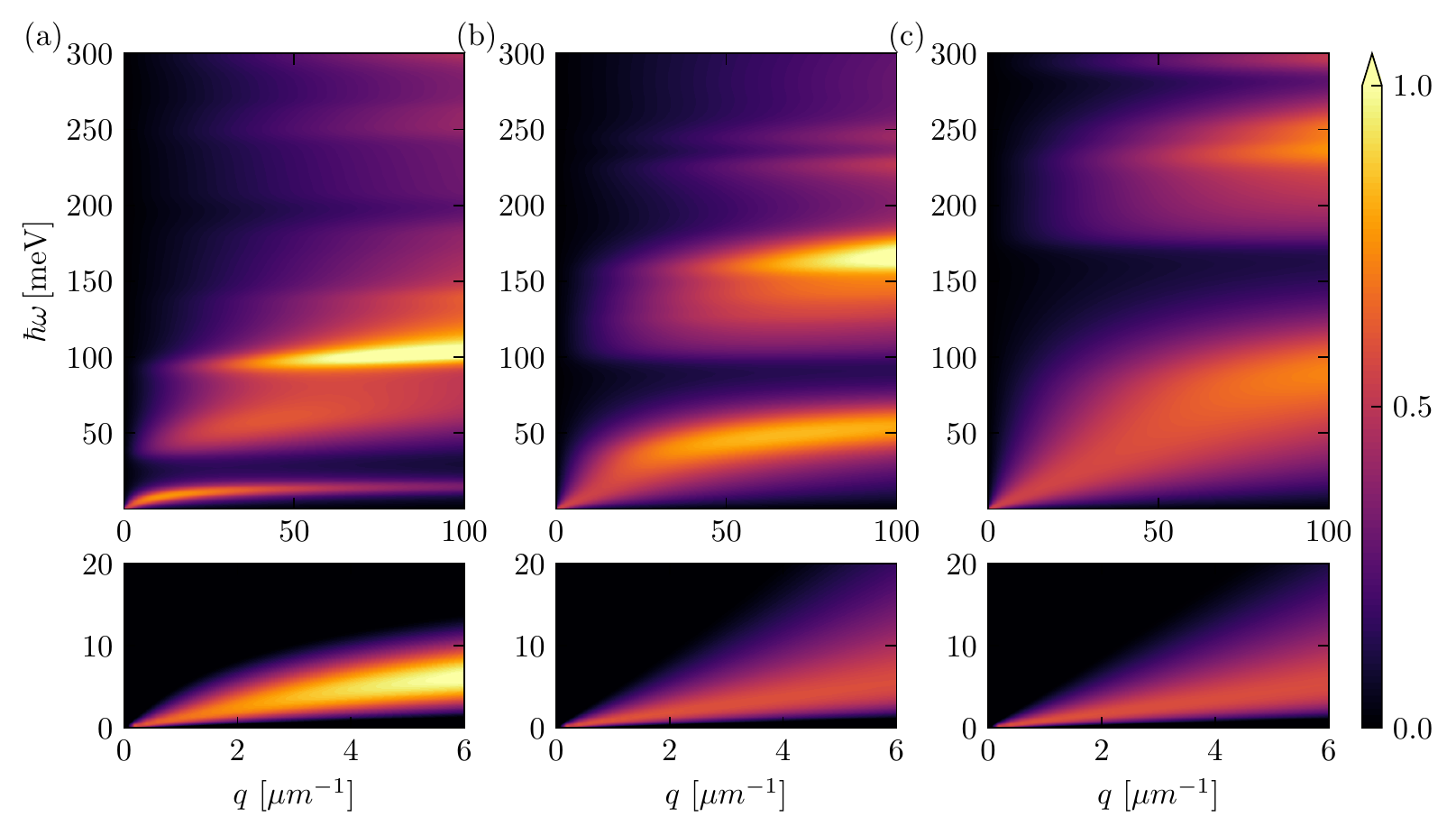}
  \caption{\label{fig:loss_var_theta} (Color online)  2D plots of the loss function $\mathcal{L}(\bm{q},\omega)$ for different values of the twist angles $\theta$ at $\xi=0$, $u_0=79.7~{\rm meV}$, $u_1 = 97.5~{\rm meV}$, $T = 5~{\rm K}$, and $\bar{\varepsilon}(0) = 4.9$. In all the panels, the lower sub panels zoom in on a smaller region of the energy-momentum plane. Data displayed in this figure have been obtained by employing the self-consistent Hartree approximation at the CNP ($\xi=0$). Panel (a) $\theta = 1.05\degree$. Panel (b) $\theta = 1.35\degree$. Panel (c) $\theta = 1.65\degree$.}
\end{figure*}
\begin{figure*}[h!]
  \includegraphics{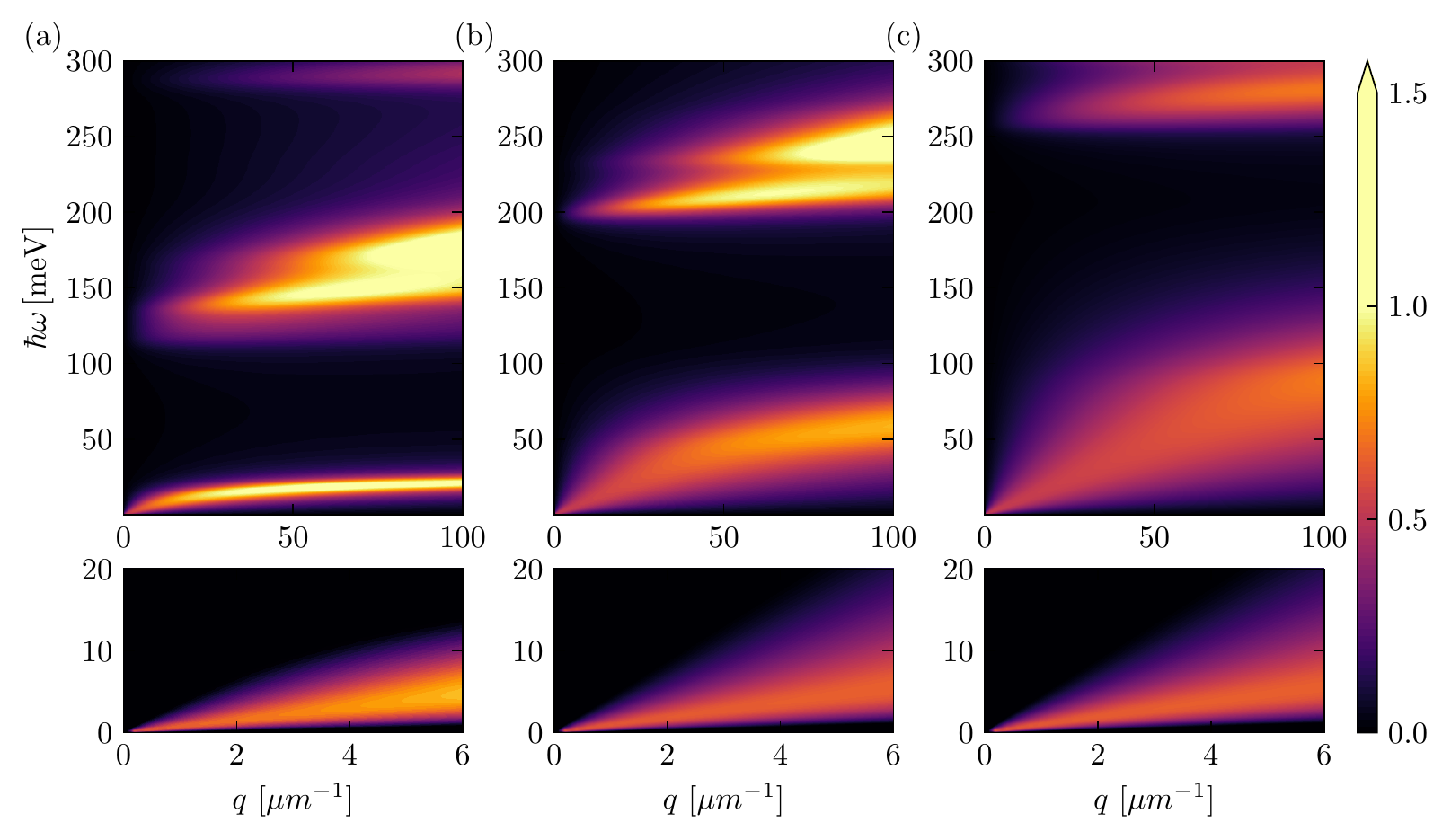}
  \caption{\label{fig:loss_var_theta_vacuum_chiral_symm} (Color online)  2D plots of the loss function $\mathcal{L}(\bm{q},\omega)$ for different values of the twist angles $\theta$ at $\xi=0$, $u_0=0~{\rm meV}$, $u_1 = 97.5~{\rm meV}$, $T = 5~{\rm K}$, and $\bar{\varepsilon}(0) = 4.9$. In all the panels, the lower sub panels zoom in on a smaller region of the energy-momentum plane. Data displayed in this figure have been obtained by employing the self-consistent Hartree approximation at the CNP ($\xi=0$). Panel (a) $\theta = 1.05\degree$. Panel (b) $\theta = 1.35\degree$. Panel (c) $\theta = 1.65\degree$.}
\end{figure*}

The loss function of TBG at three different twist angles is shown in Fig.~\ref{fig:loss_var_theta}. In its lower panels, a low-energy, low-momentum plasmon branch can be identified. Once again, this originates from inter-band transitions because, at the CNP and at $T = 5~{\rm K}$, intra-band plasmon modes are practically absent. This inter-band plasmon branch appears however as a rather broad peak in the loss function, i.e.~it is strongly damped. With the help of the upper panels, we  see that it is found at excitation energies $\hbar\omega \lesssim 20~{\rm meV}$ in the case of $\theta = 1.05\degree$. For $\theta = 1.35\degree$ and $\theta = 1.65\degree$, on the other hand, the low-energy, low-momentum branch does not extend to large momenta and energies, progressively disappearing as $q$ and $\omega$ increase. At higher energies, various peaks in the loss function can be identified. The clearest ones are: 1) one at  $\hbar\omega \approx 100~{\rm meV}$ for $\theta = 1.05\degree$, 2) one at $\hbar\omega \approx 50~{\rm meV}$ and one at $\hbar\omega \approx 170~{\rm meV}$  for $\theta = 1.35\degree$, and, finally, 3) one at $\hbar\omega \approx 250~{\rm meV}$ for $\theta = 1.65\degree$.

In Figure~\ref{fig:loss_var_theta_vacuum_chiral_symm} we show again the twist-angle dependence of the loss function, but in this case for $u_0 = 0~{\rm meV}$,
corresponding to the idealized chirally-symmetric continuum model~\cite{tarnopolsky_prl_2019}.
It is evident that reducing $u_0$ leads to a much stronger and more dispersive (and therefore propagating) inter-band plasmon mode, in the energy range $\approx 150$-$250~{\rm meV}$.
This suggests that one can gain information about the value of $u_0$ by measuring the inter-band plasmon dispersion.
The energy of the inter-band plasmon shifts towards higher energies with increasing angle, in agreement with the upward shift observed for all the optical transitions in panel (b) of Fig.~\ref{fig:optical_conductivity_var_theta}. Also its intensity seems to decrease monotonically with increasing angle.

Since plasmon modes delicately depend on $\theta$, $u_0$, and $\xi$, care needs to be exercised when color plots of the loss function referred to different sets of parameters are compared with each other. For example, as showed in Ref.~\onlinecite{hesp_arxiv_2019}, for $\theta = 1.35\degree$ and $u_0 = 0$, a clear inter-band plasmon mode emerges at energy $\hbar\omega \gtrsim 100~{\rm meV}$. 

Numerical results for the loss function of TBG encapsulated between two hexagonal Boron Nitride crystal slabs, where the frequency-dependence of $\bar{\varepsilon}(\omega)$ cannot be neglected, are reported in Appendix~\ref{app:hBN_loss}.

\section{Summary and conclusions}
\label{sec:conclusions}
We have calculated the optical conductivity and energy loss function of twisted bilayer graphene, for a wide range of microscopic parameters. In particular, we have focussed on the dependence of these properties on the  intra-sublattice inter-layer tunneling rate $u_0$ and ground-state charge density inhomogeneity. 

Away from the charge neutrality point, we have showed that the low-frequency components of the optical conductivity, i.e.~the ones governed by the Drude weight $e^2\mathcal{W}^{(0)}/\hbar^{2}$, are sensibly modified by the Hartree potential in Eq.~\eqref{eq:hartree_potential}. In particular, we found a significant enhancement of the particle-hole asymmetry of $\mathcal{W}^{(0)}$. Conversely, the high-frequency components of the optical conductivity are pretty much unaffected by the Hartree potential, and their dependence on the filling factor is also very weak. The loss function reflects all these facts. The low-frequency peaks are well described, away from the charge neutrality point, by the result in  Eq.~\eqref{eq:sqrt_q_plasmon_dispersion} and depend directly on $\mathcal{W}^{(0)}$. Conversely, the high-frequency peaks arising from inter-band transitions are virtually independent of the filling factor. 

As a byproduct of our calculations, we obtained the Seebeck coefficient in the relaxation time approximation.
Our result suggests that a strong thermoelectric effect should persist down to temperatures of $\approx 5~{\rm K}$. In the near future, therefore, photocurrent mapping techniques at cryogenic temperatures may prove to be valid tools to study the onset of the transition to broken symmetry states.

At filling factor $\xi = 0$, i.e.~at the charge neutrality point, we have evaluated $\sigma(\omega)$ and $\mathcal{L}(\bm{q},\omega)$ for different values of the intra-sublattice inter-layer tunneling energy $u_0$. The Drude weight $e^2\mathcal{W}^{(0)}/\hbar^{2}$ is a monotonically increasing function of $u_0$, which is practically insensitive to the Hartree potential~\eqref{eq:hartree_potential}. The (real part of the) inter-band contribution to the optical conductivity is not affected by the Hartree potential as well, whereas it shows a very interesting dependence on $u_0$. The position of the peaks in ${\rm Re}[\sigma^{{\rm inter}}(\omega)]$ associated to optical transitions between flat bands and neighbouring bands decreases monotonically with $u_0$. It is important to keep in mind that $u_0$ can be modified by extrinsic factors such as strain present in the samples, resulting in a sensible alteration of the optical properties of TBG. Similarly, the  peaks in the energy loss function that are related to inter-band optical transitions are strongly affected by the value of the intra-sublattice inter-layer hopping energy. This is agreement with recent experimental work~\cite{hesp_arxiv_2019}. Indeed, the authors of Ref.~\onlinecite{hesp_arxiv_2019} noted that a good match between experimental results and theory was possibile only when the value of $u_0$ used for theoretical predictions was substantially smaller than that reported in the literature~\cite{koshino_prx_2018,carr_prresearch_2019}.

Finally, we have also studied the dependence of $\sigma(\omega)$ and $\mathcal{L}(\bm{q},\omega)$ on the twist angle $\theta$, again at the charge neutrality point. We have showed that the low-frequency $\omega \simeq 0$ component of the optical conductivity, determined by $\mathcal{W}^{(0)}$, can be approximated by the value obtained for linear-dispersing energy bands only if the value of $k_{{\rm B}}T$ is much smaller than the energy range over which the valence and conduction bands are linear. This condition does not hold true in TBG with $\theta \lesssim 1.2\degree$ already at  $T = 5~{\rm K}$, showing that a description of TBG based on a linear approximation of the energy bands in not sufficient at angles close to the magic one. The (real part of the) inter-band optical conductivity has peaks at energies which increase monotonically with the twist angle.

In the future, we plan to extend our theory to include excitonic effects and to study the dependence of the static and dynamical polarization function on $u_0$ and Hartree self-consistency.

\acknowledgments

This work was supported by the European Union's Horizon 2020 research and innovation programme under grant agreements no.~785219 - GrapheneCore2 and no.~881603 - GrapheneCore3.

F.T. also acknowledges support from the SNS-WIS joint lab QUANTRA.

I.T. also acknowledges funding from the Spanish Ministry of Science, Innovation and Universities (MCIU) and State Research Agency (AEI) via the Juan de la Cierva fellowship n.~FJC2018-037098-I.

F.H.L.K. also acknowledges financial support from the Government of Catalonia through the SGR grant, and from the Spanish Ministry of Economy and Competitiveness, through the ``Severo Ochoa" Programme for Centres of Excellence in R\&D (SEV-2015-0522), support by Fundacio Cellex Barcelona, Generalitat de Catalunya through the CERCA program,  the Mineco grants Plan Nacional (FIS2016-81044-P), and the Agency for Management of University and Research Grants (AGAUR) 2017 SGR 1656.  
Furthermore, the research leading to these results has received funding from the European Union's Horizon 2020 research and innovation programme under grant agreements no. 820378 (Quantum Flagship) and no.~726001 (ERC TOPONANOP).

\onecolumngrid
\appendix
\section{Derivation of the continuum model}
\label{app:derivation_continuum_model}

In this Appendix we present a brief derivation of the continuum model~\cite{bistritzer_pnas_2011,koshino_prx_2018} we have used in this work to describe electrons roaming in the TBG moir\'e superlattice, starting from its tight-binding description. 

The basis of Bloch states used in the tight-binding description is built from the $p_z$ atomic orbitals of Carbon. These Bloch states are defined by
\begin{equation}\label{eq:bloch_states}
|\bm{k},\ell,\tau\rangle = \frac{1}{\sqrt{\mathcal{N}}}\sum_{n}e^{i(\bm{t}_{n,\ell} + \bm{d}_{\tau,\ell})\cdot\bm{k}}|n,\ell,\tau\rangle~.
\end{equation}
Here, $|n,\ell,\tau\rangle$ are localized atomic orbitals centered at the point $\bm{d}_{\tau,\ell} + \bm{t}_{n,\ell}$, i.e.
\begin{equation}
\langle\bm{r}|n,\ell,\tau\rangle = \phi(\bm{r} - \bm{d}_{\tau,\ell} - \bm{t}_{n,\ell})~,
\end{equation}
where $\phi(\bm{r})$ is the wavefunction of a $p_z$ orbital centered at the origin.
The atomic orbitals are assumed to be orthogonalized according to 
\begin{equation}
 \langle n,\ell,\tau |n^\prime,\ell^\prime,\tau^\prime\rangle= \delta_{n,n^{\prime}}\delta_{\ell,\ell^{\prime}}\delta_{\tau,\tau^{\prime}}~.
\end{equation}
In Eq.~(\ref{eq:bloch_states}), ${\cal N}$ is the number of Carbon lattice sites in each layer, $\bm{d}_{\tau,\ell}$ is the basis vector of the sublattice $\tau$ in layer $\ell$, whereas the symbol $\bm{t}_{n,\ell}$ is a shorthand for
\begin{equation}
 \bm{t}_{n,\ell} = n_{1}\tilde{\bm{t}}_{1,\ell} + n_{2}\tilde{\bm{t}}_{2,\ell} \quad \text{with } n_{1}, \, n_{2} \in \mathbb{N}~.
\end{equation}
The vectors $\tilde{\bm{t}}_{1/2,\ell}$ are primitive translation vectors of the graphene lattice in layer $\ell$,
and the sum over $n$ should be intended as
\begin{equation}
  \sum_{n} [\cdots] = \sum_{n1,n2 \in \mathbb{N}} [\cdots]~.
\end{equation}

The operators in the main text are written in the basis $|\bm{k}\rangle \otimes |\ell,\tau\rangle$, with layer and sublattice indices ordered as $\{|1A\rangle,|1B\rangle,|2A\rangle,|2B\rangle\}$.

In the two-center approximation, and retaining only the nearest-neighbour contributions, the intra-layer Hamiltonian of graphene in layer $\ell$ takes the form
\begin{equation}\label{eq:app_intra_layer_hamiltonian}
  \hat{H}^{(\ell)}_{{\rm intra}} = -t \sum_{\langle m,n \rangle}\sum_{\tau,\tau^{\prime}} |m,\ell,\tau\rangle\langle n, \ell, \tau^{\prime}|(1 - \delta_{\tau,\tau^{\prime}})~,
\end{equation}
with the energy $t$ being given by
\begin{equation}
  -t \equiv \int d\bm{r}~ \phi^{*}(\bm{r})V(\bm{r}-\bm{d}_{\tau,1})\phi(\bm{r} - \bm{d}_{\tau,1}) = \int d\bm{r}~ \phi^{*}(\bm{r})V(\bm{r}-\bm{d}_{\tau,2})\phi(\bm{r} - \bm{d}_{\tau,2})~,
\end{equation}
$V(\bm{r})$ being the spherically-symmetric potential of a Carbon atom centered at the origin. The sum over $\langle m,n\rangle$ runs over neighbouring orbitals, i.e. the states $|m,\ell,\tau\rangle$ and $|n,\ell,\tau^{\prime}\rangle$ in Eq.~\eqref{eq:app_intra_layer_hamiltonian} correspond to neighbouring orbitals. At fixed layer index $\ell$, the procedure to obtain the $\bm{k}\cdot\bm{p}$ intra-layer Hamiltonian Eq.~\eqref{eq:intralayer_hamiltonian} from the tight-binding Hamiltonian Eq.~\eqref{eq:app_intra_layer_hamiltonian} is described and explicitly carried out in Chapter 1 of Ref.~\onlinecite{katsnelson_book}, to which we refer. The core of this procedure is the calculation of the Taylor expansion of the matrix elements $\langle\bm{k},\ell,\tau|\hat{H}^{(\ell)}_{{\rm intra}}|\bm{k},\ell,\tau^{\prime}\rangle$ around $\bm{k} \approx \bm{k}_{{\rm D}}^\ell$, with $\bm{k}_{{\rm D}}^\ell$ being the wave vector at which the Dirac cone of layer $\ell$ is centered. To obtain the matrix elements $\langle\bm{k},\ell,\tau|\hat{H}^{(\ell)}_{{\rm intra}}|\bm{k},\ell,\tau^{\prime}\rangle$ explicitly, one has to choose the vectors $\tilde{\bm{t}}_{1/2,\ell}$ and $\bm{d}_{\tau,\ell}$. In this Article we have chosen the primitive translation vectors
\begin{equation}
\tilde{\bm{t}}_{1/2,\ell} = R_\ell(\theta/2)\left(\mp\frac{a}{2}, \frac{a\sqrt{3}}{2}\right)~,
\end{equation}
where $R_\ell(\theta/2)$ is defined in Eq.~(\ref{eq:rotation_matrix}).
In addition, the basis vectors are
\begin{equation}\label{eq:AB-stacking}
    \bm{d}_{\tau,\ell} =
    \begin{cases}
        \frac{a}{\sqrt{3}}  R_1(\theta/2)\left( -\frac{\sqrt{3}}{2},\frac{1}{2}\right)~,& \text{if layer} = 1 \text{ and sub-lattice} = B~.\\
        - \frac{a}{\sqrt{3}}  R_2(\theta/2)\left( -\frac{\sqrt{3}}{2},\frac{1}{2}\right)~,& \text{if layer} = 2 \text{ and sub-lattice} = A~.\\
        \bm{0}~,              & \text{otherwise~.}
    \end{cases}
\end{equation}
The choice of these translation and basis vectors is such that in the limit $\theta\rightarrow 0 $ one obtains AB-stacked bilayer graphene. With this choice, a straightforward calculation leads to
\begin{equation}
	\langle\bm{k},\ell,\tau|\hat{H}^{(\ell)}_{{\rm intra}}|\bm{k},\ell,\tau^{\prime}\rangle = te^{i\bm{k}\cdot (\bm{d}_{\tau^{\prime},\ell} - \bm{d}_{\tau,\ell})}\left[1 + e^{-i\bm{k}\cdot\tilde{\bm{t}}_{1,\ell}} + e^{i\bm{k}\cdot (- \tilde{\bm{t}}_{1,\ell} + \tilde{\bm{t}}_{2,\ell})}\right]~.
\end{equation}

We now move on to discuss the inter-layer term, i.e.~Eq.~\eqref{eq:inter_layer_hamiltonian} in the main text. In the tight-binding framework we should describe the energy involved in the tunneling of electrons between orbitals in different layers. We require this energy to be dependent on the distance $\bm{r}$ between the two orbitals and on the sublattice index of the initial and final states ($\bm{\tau^{\prime}}$ and $\bm{\tau}$, respectively), but not on the initial and final layers. The inter-layer tunneling energy will be denoted by the symbol $h_{\tau,\tau^{\prime}}(\bm{r})$. The knowledge of an explicit form of $h_{\tau,\tau^{\prime}}(\bm{r})$ is not crucial for the following calculations, as argued below and in Ref.~\onlinecite{bistritzer_pnas_2011}. In practice, such explicit form of $h_{\tau,\tau^{\prime}}(\bm{r})$ can be obtained by approximating the transfer integrals between different orbitals $\phi(\bm{r})$ mediated by the spherically-symmetric atomic potential $V(\bm{r})$. An empirical approximation in the Slater-Koster form can be found in Ref.~\onlinecite{koshino_prx_2018}. In the following, however, we will not use an explicit form of $h_{\tau,\tau^{\prime}}(\bm{r})$. 

Given the previous discussion, the inter-layer hopping term can be written as
\begin{equation}
  \hat{H}_{{\rm inter}}^{(\ell,\ell^{\prime})} \equiv \sum_{n, n^{\prime}}\sum_{\tau,\tau^{\prime}} h_{\tau,\tau^{\prime}}(\bm{d}_{\tau,\ell} + \bm{t}_{n,\ell} - \bm{d}_{\tau^{\prime},\ell^{\prime}} - \bm{t}_{n^{\prime},\ell^{\prime}})|n,\ell,\tau\rangle \langle n^{\prime},\ell^{\prime},\tau^{\prime}| + {\rm H.c.}~.
\end{equation}

Introducing the Fourier tranform of the inter-layer tunneling energy,
\begin{subequations}
\begin{equation}
h_{\tau,\tau^{\prime}}(\bm{q}) \equiv \int_{A} d\bm{r} e^{-i\bm{q}\cdot\bm{r}}h_{\tau,\tau^{\prime}}(\bm{r})~,
\end{equation}
\begin{equation}
h_{\tau,\tau^{\prime}}(\bm{r}) = \frac{1}{\mathcal{N}\Omega_{{\rm u.c.}}}\sum_{\bm{q}} e^{i\bm{q}\cdot\bm{r}}h_{\tau,\tau^{\prime}}(\bm{q})~,
\end{equation}
and the well known identity,
\begin{equation}
\frac{1}{\mathcal{N}}\sum_{n}e^{i(\bm{q}-\bm{k})\cdot\bm{t}_{n}} = \sum_{\bm{G}}\delta_{\bm{q} - \bm{k},\bm{G}}~,
 \end{equation}
\end{subequations}
we can express the matrix element $\langle \bm{p},\ell,\tau|\hat{H}_{{\rm inter}}|\bm{k},\ell^{\prime},\tau^{\prime}\rangle$ in the following form:
\begin{widetext}
 \begin{equation}\label{eq:inter_layer_hamiltonian_abstract}
 \begin{split}
  \langle \bm{p},\ell,\tau|\hat{H}_{{\rm inter}}|\bm{k},\ell^{\prime},\tau^{\prime}\rangle &= \frac{1}{\mathcal{N}^{2}\Omega_{{\rm u.c.}}}\sum_{n, n^{\prime}}\sum_{\bm{q}} e^{i(\bm{q}-\bm{p})\cdot(\bm{t}_{n,\ell} + \bm{d}_{\tau,\ell})}e^{i(\bm{k}-\bm{q})\cdot(\bm{t}_{n^{\prime},\ell^{\prime}} + \bm{d}_{\tau^{\prime},\ell^{\prime}})}h_{\tau,\tau^{\prime}}(\bm{q}) =\\&=
  \frac{1}{\Omega_{{\rm u.c.}}}\sum_{\bm{q}}\sum_{\bm{G}^{\ell}}\sum_{\bm{G}^{\ell^{\prime}}}\delta_{\bm{q} - \bm{p},\bm{G}^{\ell}}\delta_{\bm{k} - \bm{q},\bm{G}^{\ell^{\prime}}}e^{i(\bm{q}-\bm{p})\cdot\bm{d}_{\tau,\ell}}e^{i(\bm{k}-\bm{q})\cdot\bm{d}_{\tau^{\prime},\ell^{\prime}}}h_{\tau,\tau^{\prime}}(\bm{q}) =\\&=
  \frac{1}{\Omega_{{\rm u.c.}}}\sum_{\bm{G}^{\ell}}\sum_{\bm{G}^{\ell^{\prime}}}\delta_{\bm{k} - \bm{G}^{\ell^{\prime}}, \bm{p} + \bm{G}^{\ell}}e^{i\bm{G}^{\ell}\cdot\bm{d}_{\tau,\ell}}e^{i\bm{G}^{\ell^{\prime}}\cdot\bm{d}_{\tau^{\prime},\ell^{\prime}}}h_{\tau, \tau^{\prime}}(\bm{p} + \bm{G}^{\ell}) =\\&=
  \frac{1}{\Omega_{{\rm u.c.}}}\sum_{\bm{G}^{\ell}}\sum_{\bm{G}^{\ell^{\prime}}}\delta_{\bm{k} + \bm{G}^{\ell^{\prime}}, \bm{p} + \bm{G}^{\ell}}e^{i(\bm{G}^{\ell}\cdot\bm{d}_{\tau,\ell} - \bm{G}^{\ell^{\prime}}\cdot\bm{d}_{\tau^{\prime},\ell^{\prime}})}h_{\tau,\tau^{\prime}}(\bm{p} + \bm{G}^{\ell})~.
 \end{split} 
 \end{equation}
\end{widetext}
\onecolumngrid
The inter-layer Hamiltonian in the Bloch basis is thus expressed as a sum over the reciprocal lattice vectors of monolayer graphene $\bm{G}^\ell$ and $\bm{G}^{\ell^{\prime}}$ of a phase factor $e^{i(\bm{G}^{\ell}\cdot\bm{d}_{\tau,\ell} - \bm{G}^{\ell^{\prime}}\cdot\bm{d}_{\tau^{\prime},\ell^{\prime}})}$ multiplied by the Fourier transform of the inter-layer potential $h_{\tau,\tau^{\prime}}(\bm{p} + \bm{G}^{\ell})$. A drastic simplification can be performed\cite{bistritzer_pnas_2011}, by truncating the (infinite) sums over reciprocal lattice vectors in the previous equations. The truncation is justified as long as the inter-layer potential $h_{\tau,\tau^{\prime}}(\bm{p} + \bm{G}^{\ell})$ is small enough. In practice it is possible to show\cite{bistritzer_pnas_2011} that for a low-energy description of TBG, only a very small number of reciprocal lattice vectors can be retained. The vectors which one ought to retain depend on the initial choice of $\tilde{\bm{t}}_{1/2,\ell}$ and $\bm{d}_{\tau,\ell}$. Once the vectors $\bm{G}^\ell$ and $\bm{G}^{\ell^{\prime}}$ to retain are known, a simple substitution onto Eq.~\eqref{eq:inter_layer_hamiltonian_abstract} gives the phase factors $e^{i(\bm{G}^{\ell}\cdot\bm{d}_{\tau,\ell} - \bm{G}^{\ell^{\prime}}\cdot\bm{d}_{\tau^{\prime},\ell^{\prime}})}$ and values of $h_{\tau,\tau^{\prime}}(\bm{p} + \bm{G}^{\ell})$ as in Eq.~\eqref{eq:inter_layer_hamiltonian} of the main text. The last piece of information needed is the analytical form of the inter-layer tunneling potential $h_{\tau,\tau^{\prime}}(\bm{p} + \bm{G}^{\ell})$. It turns out, as anticipated above, that this is actually quite irrelevant. Since the continuum model is an approximation around $\bm{p} \approx \bm{k}_{{\rm D}}^\ell$, one can make the identification $h_{\tau,\tau^{\prime}}(\bm{k}_{{\rm D}}^\ell + \bm{G}^{\ell}) \approx h_{\tau,\tau^{\prime}}(\bm{p} + \bm{G}^{\ell})$, so that the values $u_1$ and $u_0$ discussed in the main text are just
\begin{subequations}\label{eqs:interlayer_tunneling_rates}
 \begin{equation}
    u_{0} \equiv h_{\tau,\tau}(\bm{k}_{{\rm D}}^\ell + \bm{G}^{\ell}) = h_{\tau,\tau}( \bm{k}_{{\rm D}}^\ell),
 \end{equation}
 \begin{equation}
    u_{1} \equiv h_{\tau,\tau^{\prime}}( \bm{k}_{{\rm D}}^\ell + \bm{G}^{\ell}) = h_{\tau,\tau^{\prime}}( \bm{k}_{{\rm D}}^\ell) \quad \tau \neq \tau^{\prime},
 \end{equation}
\end{subequations}
where the second equality in both of the previous equations holds true because the vectors $\bm{G}^\ell$ retained in the summation satisfy that property. This implies that instead of the full analytical form of $h_{\tau,\tau^{\prime}}(\bm{r})$ one just needs a tiny number of characteristic energy scales. These can be obtained both through tight-binding approximations~\cite{koshino_prx_2018} or density functional calculations~\cite{lucignano_prb_2019}. The inter-layer term in the main text, Eq.~\eqref{eq:inter_layer_hamiltonian} is just Eq.~\eqref{eq:inter_layer_hamiltonian_abstract} truncated to the retain only the three most relevant terms. Replacing the definitions in Eqs.~\eqref{eqs:interlayer_tunneling_rates} and \eqref{eq:AB-stacking} into the truncated sum yields directly Eq.~\eqref{eq:inter_layer_hamiltonian} of the main text.
\section{Derivation of Eq.~\eqref{eq:hartree_potential}}\label{app:hartree_potential_derivation}
The Hartree potential in the real space representation is
\begin{equation}\label{app_eq:hartree_potential}
 V^{{\rm H}}(\bm{r}) \equiv \langle \bm{r}|\hat{V}^{\rm{H}}|\bm{r}\rangle = \int d^2\bm{r}^{\prime} \frac{e^2}{\bar{\varepsilon}(0)|\bm{r} - \bm{r}^{\prime}|}n(\bm{r}^{\prime})~,
\end{equation}
where $n(\bm{r})$ is the density at position $\bm{r}$ and the integral over ${\bm r}^\prime$ is performed over the whole 2D electron system area. The Fourier expansion of the Coulomb interaction reads as following
\begin{equation}\label{app_eq:coulomb_potential}
  \frac{e^2}{\bar{\varepsilon}(0)|\bm{r} - \bm{r}^{\prime}|} = \frac{1}{A}\sum_{\bm{q}}v_{\bm q} e^{i\bm{q}\cdot(\bm{r} - \bm{r}^{\prime})}~,
\end{equation}
where
\begin{equation}\label{app_eq:coulomb_potential_fourier}
v_{\bm q} = \frac{2\pi e^{2}}{\bar{\varepsilon}(0)|\bm{q}|} = e^{2} L_{{\bm q}, \omega=0}~.
\end{equation}
As discussed previously, the eigenstates of TBG are Bloch waves
\begin{equation}
\psi_{\nu}(\bm{r},\bm{k}) = \langle \bm{r}|\bm{k}\nu\rangle = \frac{1}{\sqrt{A}}\sum_{\bm{G}}u_{\bm{G}}(\bm{k},\nu)e^{i(\bm{k} + \bm{G})\cdot\bm{r}}~.
\end{equation}
The density at position $\bm{r}$ is just the sum over the occupied states, namely 
\begin{equation}\label{app_eq:electronic_density}
\begin{split}
n(\bm{r}) &= \sum_{\bm{k}\nu}f_{\bm{k}\nu} \psi^{\dagger}_{\nu}(\bm{r},\bm{k})\psi_{\nu}(\bm{r},\bm{k}) 
= \frac{1}{A} \sum_{\bm{k}\nu}f_{\bm{k}\nu}\sum_{\bm{G}, \bm{G}^{\prime}} u^{\dagger}_{\bm{G}}(\bm{k},\nu)u_{\bm{G}^{\prime}}(\bm{k},\nu)e^{i(\bm{G}^{\prime} - \bm{G})\cdot\bm{r}} 
 \\ &= \frac{1}{A} \sum_{\bm{k}\nu}f_{\bm{k}\nu}\sum_{\bm{G}, \bm{\mathcal{G}}} u^{\dagger}_{\bm{G}}(\bm{k},\nu)u_{\bm{G} + \bm{\mathcal{G}}}(\bm{k},\nu)e^{i\bm{\mathcal{G}}\cdot\bm{r}}
\equiv \sum_{\bm{\mathcal{G}}} n(\bm{\mathcal{G}})e^{i\bm{\mathcal{G}}\cdot\bm{r}},
\end{split} 
\end{equation}
where we have introduced the quantity
\begin{equation}
  n(\bm{\mathcal{G}}) = \frac{1}{A} \sum_{\nu}\sum_{\bm{k}}f_{\bm{k}\nu}\sum_{\bm{G}} u^{\dagger}_{\bm{G}}(\bm{k},\nu)u_{\bm{G} + \bm{\mathcal{G}}}(\bm{k},\nu)~,
\end{equation}
i.e.~the Fourier component of the electron density at wave vector $\bm{\mathcal{G}}$. Now, substituting Eqs.~\eqref{app_eq:coulomb_potential},~\eqref{app_eq:coulomb_potential_fourier} and~\eqref{app_eq:electronic_density} into Eq.~\eqref{app_eq:hartree_potential}, and carrying out simple algebraic manipulations, we find
\begin{equation}\label{eq:manipulations_Hartree}
\begin{split}
V^{\rm{H}}(\bm{r}) &= \int d^2\bm{r}^{\prime} \frac{e^2}{\bar{\varepsilon}(0)|\bm{r} - \bm{r}^{\prime}|}n(\bm{r}^{\prime}) 
= \frac{1}{A}\sum_{\bm{q}}\int d\bm{r}^{\prime} v_{\bm q}e^{i\bm{q}\cdot(\bm{r} - \bm{r}^{\prime})}n(\bm{r}^{\prime})
= \frac{1}{A}\sum_{\bm{q}}\sum_{\bm{\mathcal{G}}}\int d\bm{r}^{\prime} v_{\bm q}e^{i\bm{q}\cdot(\bm{r} - \bm{r}^{\prime})} n(\bm{\mathcal{G}})e^{i\bm{\mathcal{G}}\cdot\bm{r}^{\prime}}\\
&= \sum_{\bm{q}}e^{i\bm{q}\cdot\bm{r}}\sum_{\bm{\mathcal{G}}} v_{\bm q} n(\bm{\mathcal{G}})\delta_{\bm{q}, \bm{\mathcal{G}}}
= \sum_{\bm{\mathcal{G}}} v_{\bm{\mathcal{G}}} n(\bm{\mathcal{G}})e^{i\bm{\mathcal{G}}\cdot\bm{r}}
= \sum_{\bm{\mathcal{G}}} \frac{2\pi e^{2}}{\bar{\varepsilon}(0)|\bm{\mathcal{G}}|} n(\bm{\mathcal{G}})e^{i\bm{\mathcal{G}}\cdot\bm{r}}~,
\end{split} 
\end{equation}
where we have used that
\begin{equation}
  \frac{1}{A}\int d^2\bm{r} e^{-i\bm{q}\cdot\bm{r}} = \delta_{\bm{q},\bm{0}}~.
\end{equation}
Eq.~\eqref{eq:hartree_potential} follows from (\ref{eq:manipulations_Hartree}), after recalling that
\begin{equation}
V^{{\rm H}}(\bm{r}) = \langle \bm{r}|\hat{V}^{\rm{H}}|\bm{r}\rangle~.
\end{equation}
As explained in the main text, to ensure overall charge neutrality due to the positively charged background~\cite{Giuliani_and_Vignale}, one has to exclude the term with $\bm{\mathcal{G}}={\bm0}$ from the sum in the last term of Eq.~\eqref{eq:manipulations_Hartree}.

\section{Proof of Eqs.~(\ref{eq:optical_conductivity_intra})-(\ref{eq:optical_conductivity_inter})}
\label{app:conductivity}
The electrical conductivity $\sigma_{\alpha \beta}(\bm r,\bm r',\omega)$ of an electron system is defined as the linear response function connecting the {\it electrical} current at position $\bm r$ to the {\it total} applied electric field at position $\bm r'$, i.e.
\begin{equation}\label{eq:cond_def_real_space}
J_\alpha^{\rm el}(\bm r,\omega)=\int d^D \bm r'\sigma_{\alpha \beta}(\bm r,\bm r',\omega)E_\beta^{\rm tot}(\bm r',\omega)~,
\end{equation}
where $J_\alpha^{\rm el}(\bm r,\omega)$ is the $\alpha$-th Cartersian component of the electrical current at position $\bm r$ and frequency $\omega$, $E_\beta^{\rm tot}(\bm r',\omega)$ is the $\beta$-th Cartesian component of the {\it total} applied electric field at position ${\bm r}^\prime$ and frequency $\omega$, and $d^D \bm r'$ denotes the measure of integration in $D$-dimensional space. From now on, Greek letters will denote Cartesian indices and the Einstein summation convention over repeated Greek indices is understood.

By Fourier transforming both members of Eq.~(\ref{eq:cond_def_real_space}) we obtain
\begin{equation}\label{eq:cond_def_rec_space}
J_{\bm q \alpha}^{\rm el}(\omega)=\sum_{\bm q'}\sigma_{\alpha \beta}(\bm q,\bm q',\omega)E_{\bm q'\beta}^{\rm tot}(\omega)~,
\end{equation}
where we defined 
\begin{align}\label{eq:ft_definitions}
J_{\bm q \alpha}^{\rm el}(\omega) & = \int d^D \bm r~ e^{-i \bm q \cdot \bm r} J_\alpha^{\rm el}(\bm r,\omega)~,\\
E_{\bm q \beta}^{\rm tot}(\omega) & = \int d^D \bm r~ e^{-i \bm q \cdot \bm r} E_\beta^{\rm tot}(\bm r,\omega)~,\\
\sigma_{\alpha \beta}(\bm q,\bm q',\omega) & = \frac{1}{V}\int d^D \bm r~ e^{-i \bm q \cdot \bm r}\int d \bm r' ~e^{i \bm q' \cdot \bm r'}\sigma_{\alpha \beta}(\bm r,\bm r',\omega)~,
\end{align}
and $V$ is the electron system volume in $D$ spatial dimensions.

We now consider a system of non-interacting electron of mass $m$ and charge $-e<0$, whose dynamics is controlled by the Hamiltonian (in first quantization)
\begin{equation}\label{eq:hamiltonian}
\hat{H}(t)=\sum_{i}\left[ \frac{1}{2m}\left(\hat{\bm p}_i +\frac{e}{c} \bm A_1(\hat {\bm r}_i,t)\right)^2-e\phi_0(\hat{\bm r_i})\right]~,
\end{equation}
where $\hat {\bm r}_i$, and $\hat{\bm p}_i$ are the position and momentum operators of the $i$-th electron, respectively, $\phi_0(\bm r)$ is an external, static, scalar electric potential, and $\bm A_1(\bm r,t)$ is a time-dependent vector potential perturbation. We note that any time-dependent scalar perturbation can be written as a vector potential using a gauge transformation~\cite{Giuliani_and_Vignale}.

In the spirit of linear response theory~\cite{Giuliani_and_Vignale}, we can expand the Hamiltonian with respect to the perturbation as
\begin{equation}\label{eq:hamiltonian_expansion}
\hat{H}(t) = \hat{H}_0 + \hat{H}_1(t) +\mathcal{O}( A_1^2)~,
\end{equation}
where
\begin{equation}\label{eq:hamiltonian_unperturbed}
\hat{H}_0=\sum_{i}\left[ \frac{1}{2m}\hat{\bm p}_i^2-e\phi_0(\hat{\bm r_i})\right]~,
\end{equation}
is the unperturbed Hamiltonian, and
\begin{equation}\label{eq:hamiltonian_perturbation}
\begin{split}
\hat{H}_1(t) & =\sum_{i} \frac{e}{2mc}\left\{\hat{ p}_{i,\alpha} ; A_{1\alpha}(\hat {\bm r}_i,t)\right\}
=\int d^D\bm r~A_{1\alpha}(\bm r,t)\sum_{i} \frac{e}{2mc}\left\{\hat{ p}_{i,\alpha} ; \delta(\hat {\bm r}_i-\bm r) \right\}
=\int d^D\bm r~\frac{e}{c}A_{1\alpha}(\bm r,t) \hat{J}^{\rm p }_\alpha(\bm r)
\end{split}
\end{equation}
is the perturbation Hamiltonian. Here, $\hat{J}^{\rm p }_\alpha(\bm r)$ is the {\it paramagnetic} current density operator
\begin{equation}
\hat{J}^{\rm p}_\alpha(\bm r) = \sum_{i}\left[ \frac{1}{2m}\left\{\hat{ p}_{i,\alpha} ; \delta(\hat {\bm r}_i-\bm r) \right\} \right]~.
\end{equation}
The physical {\it particle} current density operator at a position $\bm r$ is instead given by
\begin{equation}\label{eq:current_operator}
\hat{J}_{\alpha}(\bm r)
=\sum_i \left[\frac{1}{2m} \left\{\hat{p}_{i,\alpha}+\frac{e}{c} A_{1,\alpha}(\bm r,t);\delta(\hat {\bm r}_i-\bm r) \right\}\right] 
= \hat{J}^{\rm p}_\alpha(\bm r)+\frac{e}{mc} A_{1,\alpha}(\bm r,t)\hat{n}(\bm r)~,
\end{equation}
where the particle density operator is given by
\begin{equation}\label{eq:density_operator}
\hat{n}(\bm r) = \sum_{i} \delta(\hat {\bm r}_i-\bm r)~.
\end{equation}
The expectation value of the current operator is therefore 
\begin{equation}
\begin{split}
J_\alpha(\bm r,t) & \equiv \tr[\hat{\rho}(t)\hat{J}_{\alpha}(\bm r)] =
\tr[\hat{\rho}_0\hat{J}^{\rm p}_{\alpha}(\bm r)] 
+ \int_0^\infty d\tau \int d^D\bm r' \frac{e}{c}A_{1\beta}(\bm r',t-\tau) \chi_{ \hat{J}^{\rm p }_\alpha(\bm r), \hat{J}^{\rm p }_\beta(\bm r')}(\tau)
+\frac{e}{mc} A_{1,\alpha}(\bm r,t)\tr[\hat{\rho}_0\hat{n}(\bm r)]\\
& = \int_0^\infty d\tau \int d^D\bm r' \frac{e}{c}A_{1\beta}(\bm r',t-\tau) \chi_{ \hat{J}^{\rm p }_\alpha(\bm r), \hat{J}^{\rm p }_\beta(\bm r')}(\tau)
+\frac{e}{mc} A_{1,\alpha}(\bm r,t) n(\bm r).
\end{split}
\end{equation}
Here $\hat{\rho}(t)$ is the density operator of the many-body system, $\hat{\rho}_0$ is its equilibrium value, and we used the notation of Ref.~\onlinecite{Giuliani_and_Vignale}. 

By Fourier transforming with respect to space and time, making use of $J_{\bm q \alpha}^{\rm el}(\omega) = -eJ_{\bm q \alpha}(\omega) $, and $\bm A=-(ic/\omega) \bm E $, and comparing with (\ref{eq:cond_def_rec_space}), we finally find:
\begin{equation}\label{eq:sigma}
\sigma_{\alpha\beta}(\bm q,\bm q',\omega)=\frac{ie^2}{\omega V}\left[\chi_{\hat{J}_{\bm q \alpha}^{\rm p}\hat{J}_{-\bm q' \beta}^{\rm p}}(\omega)+ \frac{\delta_{\alpha\beta}}{m}\langle\hat{n}_{\bm q-\bm q'}\rangle\right]~,
\end{equation}
where the Fourier transforms of the density and current density operators are give by, respectively,
\begin{equation}\label{eq:density_reciprocal}
\hat{n}_{\bm q} = \sum_{i} e^{-i\bm q \cdot \hat {\bm r}_i}
\end{equation}
and
\begin{equation}\label{eq:current_reciprocal}
\hat{J}^{\rm p }_{\bm q\alpha} = \sum_{i} \frac{1}{2m}\left\{\hat{ p}_{i,\alpha} ; e^{-i\bm q \cdot \hat {\bm r}_i} \right\}
=\sum_{i}\frac{1}{2m}\left\{\hat{ p}_{i,\alpha} ; \hat{n}_{\bm q} \right\}~.
\end{equation}

The paramagnetic current-current response function in Eq.~(\ref{eq:sigma}) can be written, at the non-interacting level, as 
\begin{equation}\label{eq:current_response}
\chi_{\hat{J}_{\bm q \alpha}^{\rm p}\hat{J}_{-\bm q' \beta}^{\rm p}}(\omega) = \sum_{m n}\frac{f_m-f_n}{\epsilon_m-\epsilon_n+\hbar\omega+i\eta}\langle m|\hat{J}_{\bm q \alpha}^{\rm p}|n\rangle\langle n|\hat{J}_{-\bm q' \beta}^{\rm p}|m \rangle~,
\end{equation}
where $\left\{|m\rangle\right\}$ is a complete set of eigenstates of $\hat{H}_0$ and $\epsilon_m$ are the corresponding energies.
In a crystal, Bloch translational invariance implies that the wave vectors $\bm q$ and $\bm q'$ can differ at most by a reciprocal lattice vector:
the conductivity can therefore be written as 
\begin{equation}\label{eq:conductivity_non_local_crystal}
\sigma_{\alpha\beta}^{\bm G \bm G'}(\bm q, \omega) \equiv \sigma_{\alpha\beta}(\bm q + \bm G,\bm q + \bm G',\omega)~, 
\end{equation}
with $\bm q$ in the first Brillouin zone and $\bm G$, $\bm G'$ reciprocal lattice vectors.
Choosing a base of eigenstates on $\hat{H}_0$ in the Bloch form $|\bm k \nu\rangle$, Eq.~(\ref{eq:sigma}) can be recast in the form
\begin{equation}\label{eq:sigma2}
\begin{split}
\sigma_{\alpha\beta}^{\bm G {\bm G}^\prime}(\bm q,\omega) = & \frac{ige^2\hbar}{V}\sum_{\bm k, \nu, \nu^\prime}
\left[-\frac{f_{\bm k \nu}-f_{\bm k + \bm q \nu^\prime}}
{\epsilon_{\bm k\nu}-\epsilon_{\bm k + \bm q \nu^\prime}} \right]
\frac{\langle \bm k\nu|\hat{J}_{\bm q +\bm G \alpha}^{\rm p}|\bm k + \bm q \nu'\rangle\langle \bm k + \bm q \nu'|\hat{J}_{-\bm q -\bm G' \beta}^{\rm p}|\bm k\nu \rangle}
{\epsilon_{\bm k\nu}-\epsilon_{\bm k + \bm q \nu'}+\hbar\omega+i\eta}
+\frac{ie^2}{V m \omega}T_{\alpha\beta}^{\bm G,\bm G'}(\bm q)~,
\end{split}
\end{equation}
where
\begin{equation}\label{eq:T_alphabeta}
T_{\alpha\beta}^{\bm G \bm G'}(\bm q)=mg\sum_{\bm k, \nu, \nu'}\frac{f_{\bm k \nu}-f_{\bm k + \bm q \nu^\prime}}
{\epsilon_{\bm k\nu}-\epsilon_{\bm k + \bm q \nu^\prime}} 
\langle \bm k\nu|\hat{J}_{\bm q +\bm G \alpha}^{\rm p}|\bm k + \bm q \nu'\rangle\langle \bm k + \bm q \nu'|\hat{J}_{-\bm q -\bm G' \beta}^{\rm p}|\bm k\nu \rangle
+\delta_{\alpha\beta}\langle \hat{n}_{\bm G-\bm G'}\rangle~.
\end{equation}
To find Eqs.~(\ref{eq:sigma2})-(\ref{eq:T_alphabeta})  we used the following mathematical identity:
\begin{equation}\label{eq:denominator_identity}
\frac{1}{\epsilon_m-\epsilon_n+\hbar\omega+i\eta} = \frac{1}{\epsilon_m-\epsilon_n} \left[1- \frac{\hbar\omega+i\eta}{\epsilon_m-\epsilon_n+\hbar\omega+i\eta}\right] \overset{\eta \to 0}{=} \frac{1}{\epsilon_m-\epsilon_n} \left[1- \frac{\hbar\omega}{\epsilon_m-\epsilon_n+\hbar\omega+i\eta}\right].
\end{equation}
Using that
\begin{equation}
\hat{J}_{\bm 0 \alpha}^{(0)} = \frac{1}{m} \hat{p}_{\alpha}=\frac{i}{\hbar} \left[\hat{H}_0;\hat{r}_\alpha\right]
\end{equation} 
and the canonical commutator $[\hat{r}_{ \alpha};\hat{p}_{ \beta}^{(0)}] = i \hbar\delta_{\alpha \beta}$, one can show that
\begin{equation}\label{eq:tensor_local}
\lim_{\bm q \to \bm 0}T^{\bm 0\bm 0}_{\alpha\beta}(\bm q)=0~. 
\end{equation}
The local conductivity, defined as
\begin{equation}\label{eq:local_limit}
\sigma_{\alpha\beta}(\omega) \equiv \lim_{\bm q \to \bm 0}\sigma_{\alpha\beta}^{\bm 0\bm 0}(\bm q, \omega)~,
\end{equation}
can be therefore expressed as:
\begin{equation}\label{eq:local_cond_crystal}
\begin{split}
\sigma_{\alpha \beta}(\omega) = & \frac{ie^2 g}{\hbar} \sum_{\nu}\int\frac{d^D\bm k}{(2\pi)^D}[-f_{\bm k \nu}']\frac{\langle \bm k \nu|\frac{\hbar}{m}\hat{p}_{\alpha}|\bm k \nu\rangle\langle \bm k \nu|\frac{\hbar}{m}\hat{p}_{\beta}|\bm k \nu \rangle}{\hbar\omega+i\eta}\\
+ &\frac{ie^2 g}{\hbar} \sum_{\nu\neq \nu'}\int\frac{d^D\bm k}{(2\pi)^D}\left[-\frac{f_{\bm k \nu}-f_{\bm k \nu'}}{\epsilon_{\bm k \nu}-\epsilon_{\bm k \nu'}}\right]\frac{\langle \bm k \nu|\frac{\hbar}{m}\hat{p}_{\alpha}|\bm k \nu'\rangle\langle \bm k \nu'|\frac{\hbar}{m}\hat{p}_{\beta}|\bm k \nu \rangle}{\epsilon_{\bm k \nu}-\epsilon_{\bm k \nu'}+\hbar\omega+i\eta}~.
\end{split}
\end{equation}
Here, we separated the terms of the sum with $\nu = \nu'$ and used the limit $\lim_{\bm q \to \bm 0} \frac{f_{\bm k \nu}-f_{\bm k + \bm q \nu}}
{\epsilon_{\bm k\nu}-\epsilon_{\bm k + \bm q \nu}} =f'_{\bm k\nu}$.
The matrix elements appearing in Eq.~(\ref{eq:local_cond_crystal}) can be conveniently expressed as
\begin{equation}\label{eq:kp_identity}
\langle \bm k \nu|\frac{\hbar}{m}\hat{p}_{\alpha}|\bm k \nu'\rangle = \langle u_{\bm k \nu}|\frac{\hbar}{m}[\hat{p}_{\alpha}+\hbar k_\alpha]|u_{\bm k \nu'}\rangle = \langle u_{\bm k \nu}|\partial_{\bm k_\alpha}\hat{H}(\bm k)|u_{\bm k \nu'}\rangle~,
\end{equation}
where $|u_{\bm k \nu}\rangle$ are the periodic parts of the Bloch wavefunctions and $\hat{H}(\bm k)\equiv e^{-i\bm k \cdot \bm r}
\hat{H}_0e^{i\bm k \cdot \bm r}$. This yields Eqs.~(\ref{eq:optical_conductivity_intra})-(\ref{eq:optical_conductivity_inter}) in the main text.

\section{Proof of Eq.~\eqref{eq:non_local_epsilon}}
\label{app:non_local_dielectric_function}

In a generic, not-translationally-invariant, electronic system the dielectric function relates the externally applied electric potential with the total electric potential (i.e. ~the sum of the external potential and the Hartree potential)
\begin{equation}
-e\phi_{\rm ext} (\bm q, \omega)= -e\sum_{\bm q'}\epsilon(\bm q,\bm q',\omega) \phi_{\rm tot} (\bm q', \omega)~.
\end{equation}

The dielectric function $\epsilon(\bm q,\bm q',\omega)$ can be related~\cite{Giuliani_and_Vignale,torre_prb_2017} to the {\it proper} density-density response function $\tilde{\chi}(\bm q, \bm q',\omega)$,
\begin{equation}
\epsilon(\bm q,\bm q',\omega) =  \delta_{\bm q\bm q'}-e^2L_{\bm q,\omega}\tilde{\chi}(\bm q, \bm q',\omega)~,
\end{equation}
where we have assumed that the interaction potential does not couple different wave vectors (i.e.~the dielectric environment, which alters the e-e interaction in vacuum, has translational invariance).
The proper density-density response function relates the charge density $\rho(\bm q,\omega)$ to the external potential via,
\begin{equation}
\rho(\bm q,\omega)=e^2\sum_{\bm q' }\tilde{\chi}(\bm q, \bm q',\omega)\phi_{\rm ext} (\bm q', \omega)~.
\end{equation}
Using the continuity equation, $i\bm q \cdot \bm J_{\rm el}(\bm q,\omega)-i\omega \rho(\bm q,\omega) = 0$ and $\bm E_{\rm ext} = -i\bm q \phi_{\rm ext} (\bm q, \omega)$ in Eq.~\eqref{eq:cond_def_rec_space} we obtain the following relationwhip between $\tilde{\chi}(\bm q, \bm q',\omega)$ and $\sigma_{\alpha\beta}(\bm q,\bm q',\omega)$:
\begin{equation}
\tilde{\chi}(\bm q, \bm q',\omega) = \frac{-i}{e^2\omega} q_\alpha q_\beta' \sigma_{\alpha\beta}(\bm q,\bm q',\omega)~.
\end{equation}

In a crystal, all the response functions can connect wave vectors that differ at most by a reciprocal lattice vector. We can therefore define
\begin{equation}
\begin{split}
\epsilon_{\bm G \bm G'}(\bm q,\omega)& \equiv \epsilon(\bm q +\bm G, \bm q + \bm G',\omega)=
 \delta_{\bm q+\bm G\,\bm q+\bm G'}-e^2L_{\bm q+\bm G,\omega}\tilde{\chi}(\bm q+\bm G, \bm q+\bm G',\omega)=\\
 & =\delta_{\bm G \bm G'} + L_{\bm q+\bm G,\omega}\frac{i(\bm q+\bm G)_\alpha(\bm q+\bm G')_\beta\sigma_{\alpha\beta}(\bm q+ \bm G,\bm q+ \bm G' ,\omega)}{\omega}~,
\end{split}
\end{equation}
where $\bm q$ lies in the first Brillouin zone and $\bm G$, $\bm G'$ are reciprocal lattice vectors. Using Eq.~\eqref{eq:conductivity_non_local_crystal} in the previous equation we immediately get Eq.~\eqref{eq:non_local_epsilon} in the main text.
\section{Loss function of TBG encapsulated in hexagonal Boron Nitride}
\label{app:hBN_loss}
In a 2D system sandwiched between two half-spaces filled with a dielectric with a frequency-dependent permittivity $\bar{\varepsilon}(\omega)$, the interaction potential appearing in Eq.~\eqref{eq:local_loss_function} reads as following
\begin{equation}\label{eq:interaction_potential}
L_{{\bm q},\omega} = \frac{2\pi}{q\bar{\varepsilon}(\omega)}~.
\end{equation}
Since high-quality samples of TBG are always encapsulated in hBN, which is an hyperbolic uniaxial dielectric~\cite{caldwell_naturecomm_2014}, we here take
\begin{equation}\label{eq:hBN}
\bar{\varepsilon}(\omega)=\sqrt{\varepsilon_{z}(\omega)\varepsilon_{x}(\omega)}~,
\end{equation}
where $\varepsilon_{z}(\omega)$ and $\varepsilon_{x}(\omega)$ are the out-of-plane and in-plane dielectric permittivities of hBN. These have the following frequency dependence~\cite{caldwell_naturecomm_2014} ($i=x,z$)
\begin{equation}\label{eq:parametrization_permittivities}
\epsilon_{i}(\omega) = \epsilon_{i}(\infty) + \frac{s_{i}\hbar^2 \omega^2_{i}}{\hbar^2 \omega^2_{i}-i\hbar^2 \gamma_{i}\omega-\hbar^2\omega^2}~,
\end{equation}
with parameters given in Table~\ref{table_parameters_hBN}. Note that with the parametrization (\ref{eq:parametrization_permittivities}) of the frequency dependence of the permittivities $\epsilon_{i}(\omega)$, we have $\bar{\varepsilon}(0)=4.9$, in agreement with the value used in the main text. 

In writing Eq.~(\ref{eq:hBN}) we have neglected finite-thickness effects and assumed that TBG is encapsulated between two semi-infinite hBN crystal slabs. Finite-thickness effects can be accounted for by introducing suitable $q$-dependent form factors~\cite{hesp_arxiv_2019,forcellini_naturenano_2017} in Eq.~(\ref{eq:interaction_potential}).

\begin{figure*}[h!]
  \includegraphics{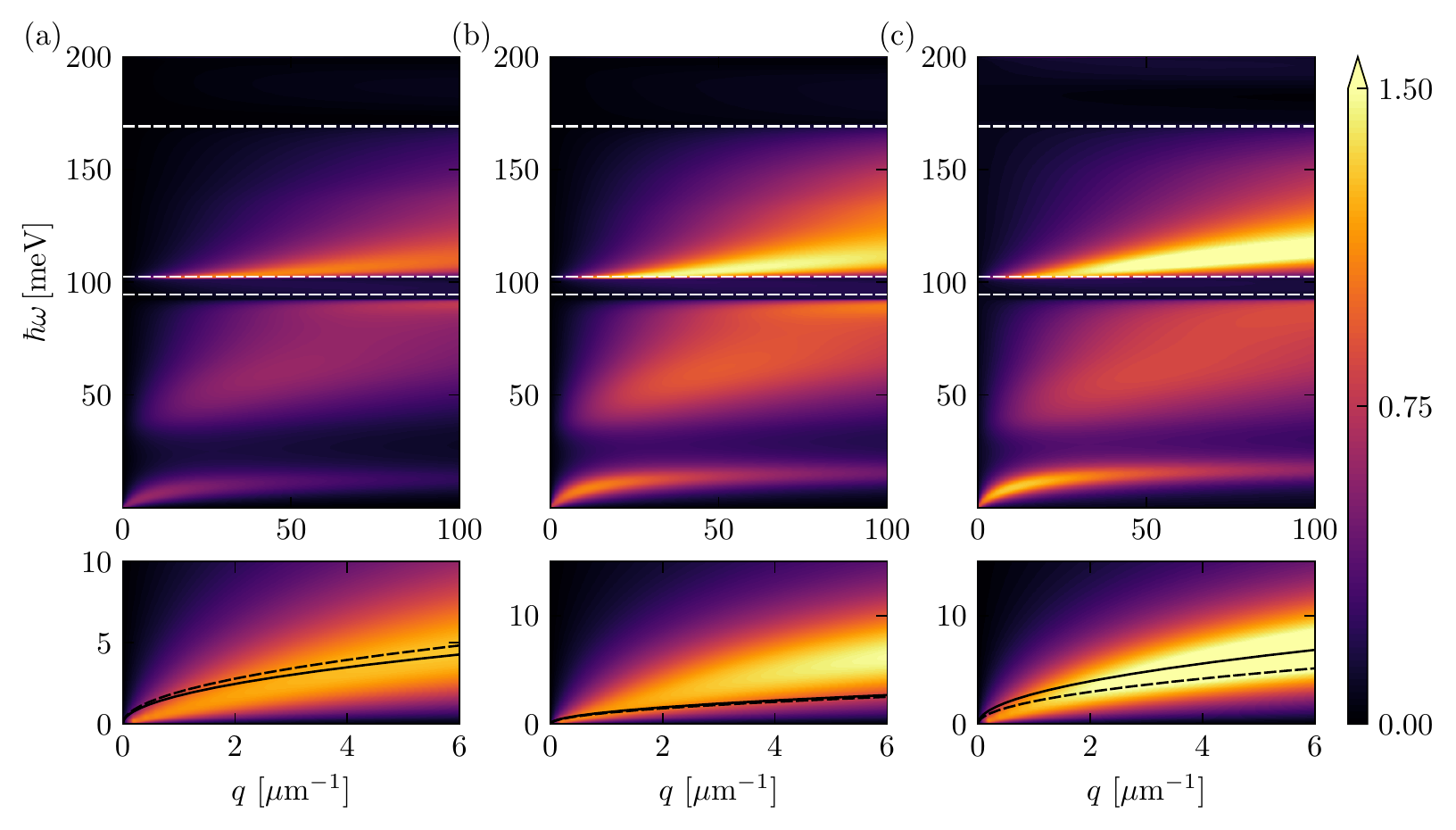}
\caption{\label{fig:loss_function_var_filling_hBN} (Color online) 2D plots of the energy loss function $\mathcal{L}(\bm{q},\omega)$ of TBG encapsulated in hBN, for different values of the filling factor $\xi$ at $\theta = 1.05\degree$, $u_0 = 79.7~{\rm meV}$, $u_1 = 97.5~{\rm meV}$, and $T = 5~{\rm K}$: (a) hole doping, $\xi = - 3/4$; (b) CNP, $\xi=0$; (c) electron doping, $\xi = +3/4$. In all the panels,  the lower sub panels zoom in on a smaller region of the energy-momentum plane. All 2D plots displayed in this figure have been obtained by using the self-consistent Hartree approximation. The black solid (dashed) lines are the analytical intra-band plasmon dispersions calculated through Eq.~\eqref{eq:sqrt_q_plasmon_dispersion}, making use of eigenvalues and eigenvectors of the self-consistent Hartree (bare) Hamiltonian in Eq.~\eqref{eq:abstract_full_hamiltonian} (Eq.~\eqref{eq:abstract_hamiltonian}), respectively. The white dashed lines denote the bounds of the hBN reststrahlen bands. The upper edge of the upper reststrahlen band is outside of the range of values of $\hbar\omega$ shown in this figure.}
\end{figure*}

\begin{figure*}[h!]
  \includegraphics{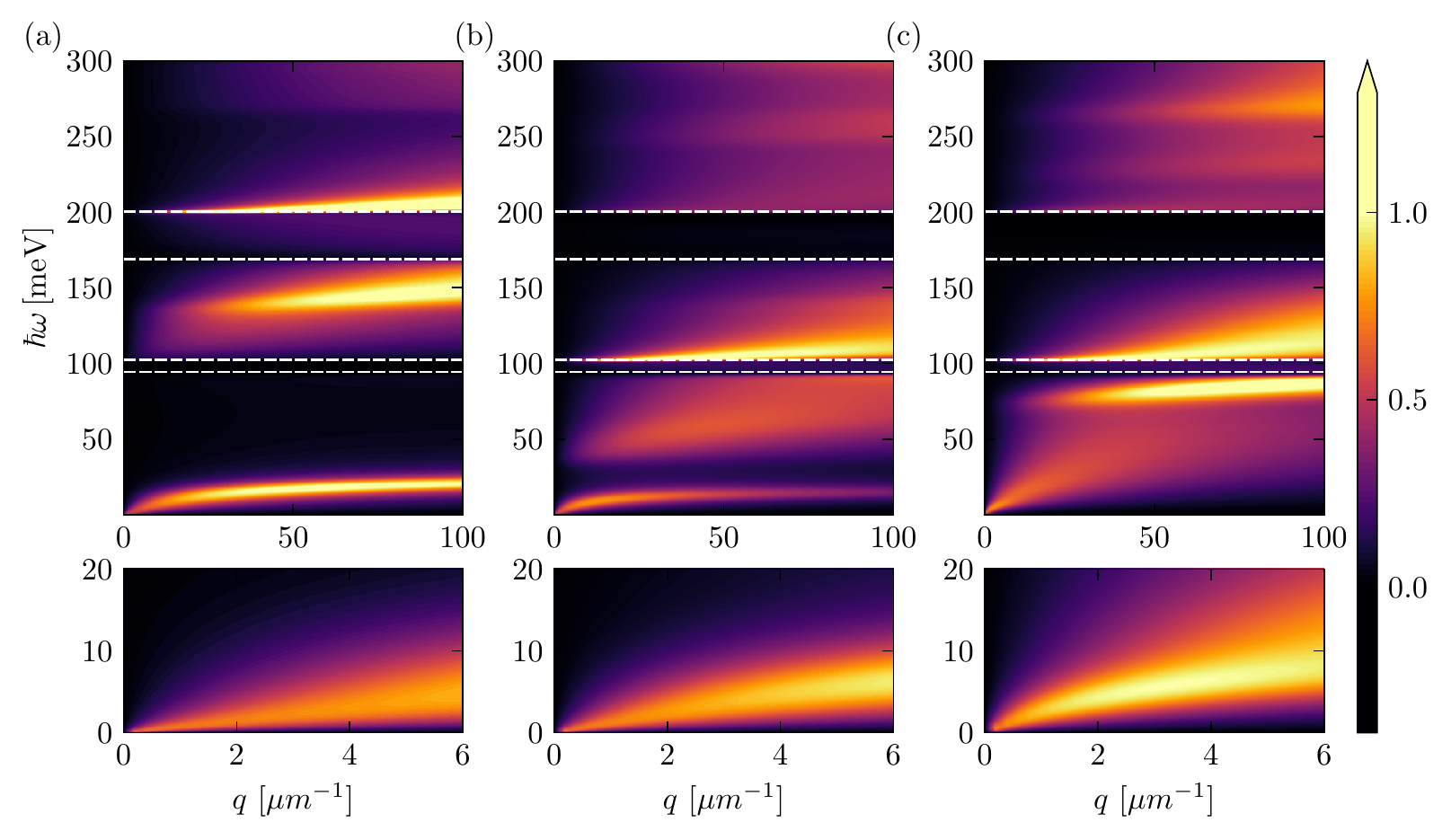}
  \caption{\label{fig:loss_var_delta_hBN} (Color online) 2D plots of the loss function $\mathcal{L}(\bm{q},\omega)$ of TBG encapsulated in hBN, for different values of the intra-sublattice hopping energy $u_0$ at $\theta = 1.05\degree$, $u_1 = 97.5~{\rm meV}$, $\xi = 0$, and $T = 5~{\rm K}$: (a) $u_0 = 0$ as in Ref.~\onlinecite{tomarken_prl_2019}; (b) $u_0 = 48.2~{\rm meV}$; (c) $u_0 = u_1 = 97.5~{\rm meV}$ as in Ref.~\onlinecite{bistritzer_pnas_2011}. In all the panels, the lower sub panels zoom in on a smaller region of the energy-momentum plane. Data displayed in this figure have been obtained by employing the self-consistent Hartree approximation at the CNP ($\xi=0$). The white dashed lines denote the bounds of the hBN reststrahlen bands.}
\end{figure*}

\begin{figure*}[h!]
  \includegraphics{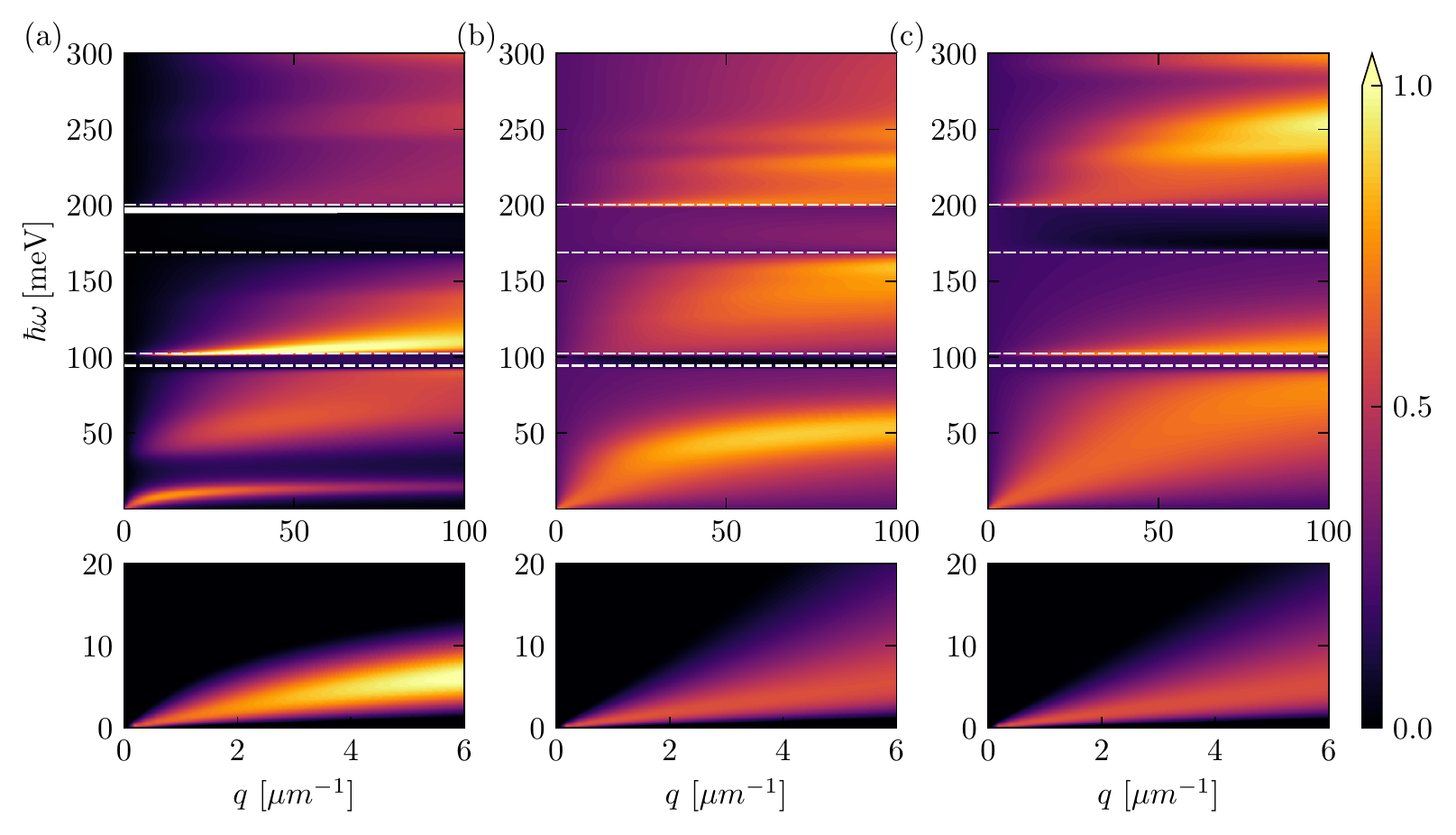}
  \caption{\label{fig:loss_var_theta_hBN}(Color online) 2D plots of the loss function $\mathcal{L}(\bm{q},\omega)$ of TBG encapsulated in hBN, for different values of the twist angles $\theta$ at $\xi=0$, $u_0=79.7~{\rm meV}$, $u_1 = 97.5~{\rm meV}$, and $T = 5~{\rm K}$. In all the panels, the lower sub panels zoom in on a smaller region of the energy-momentum plane. Data displayed in this figure have been obtained by employing the self-consistent Hartree approximation at the CNP ($\xi=0$). Panel (a) $\theta = 1.05\degree$. Panel (b) $\theta = 1.35\degree$. Panel (c) $\theta = 1.65\degree$. The white dashed lines denote the bounds of the hBN reststrahlen bands.}
\end{figure*}

\begin{table}[t]
\begin{ruledtabular}
\begin{tabular}{l | c c}
\, & $i=x$  & $i=z$	\\
\hline
$s_{i}$	& 2.001 & 0.5262 \\
$\epsilon_{i}(\infty)$ & 4.9 & 2.95 \\	
$\hbar \omega_{i}~({\rm meV})$ & 168.6 & 94.2 \\
$\hbar \gamma_{i}~({\rm meV})$ & 0.87 & 0.25\\
\end{tabular}
\end{ruledtabular}
\caption{The parameters entering the bulk hBN dielectric functions in Eq.~(\ref{eq:hBN}). These values have been extracted from Ref.~\onlinecite{caldwell_naturecomm_2014}.\label{table_parameters_hBN}}
\end{table}
The loss function of TBG encapsulated in hBN evaluated for different values of $\xi$, $u_0$ and $\theta$ is shown in Figs.~\ref{fig:loss_function_var_filling_hBN},~\ref{fig:loss_var_delta_hBN}, and \ref{fig:loss_var_theta_hBN}, respectively. As before, the upper panels display $\mathcal{L}(\bm{q},\omega)$ at energy scales which are suited to inspect collective excitations originating from inter-band processes. Inter-band plasmon branches are clearly visible at energies $\hbar\omega >  50~{\rm meV}$ in any of the three figures, and are analogous to the ones experimentally measured in Ref.~\onlinecite{hesp_arxiv_2019}. Qualitatively, the inter-band plasmons of hBN-encapsulated TBG have similar features with respect to those calculated by neglecting the frequency dependence of $\bar{\varepsilon}(\omega)$, as in the main text. For the most part, the filling factor $\xi$ leaves their position in the $\omega$-$q$ plane  unaltered. Conversely, both the inter-layer hopping amplitude and the twist angle have a higher impact on the inter-band plasmons. The inter-layer hopping amplitude, in particular, shifts the characteristic frequency of the inter-band plasmon from $\hbar\omega \approx 140~{\rm meV}$ at $u_{0} =0$ down to $\hbar\omega \approx 80~{\rm meV}$ at $u_{0} = 97.5~{\rm meV}$. 

In Figs.~\ref{fig:loss_function_var_filling_hBN},~\ref{fig:loss_var_delta_hBN} and~\ref{fig:loss_var_theta_hBN} we have clearly highlighted the hBN reststrahlen bands in the energy intervals $94 ~{\rm meV}\leq\hbar \omega \leq102 ~{\rm meV}$ (lower reststrahlen band) and $170 ~{\rm meV}\leq\hbar \omega \leq 200 ~{\rm meV}$ (upper reststrahlen band). These bounds can be easily found by looking at the (four) frequencies at which the product $\varepsilon_x(\omega)\varepsilon_z(\omega)$ changes sign. Inside the reststrahlen bands $\varepsilon_x(\omega)\varepsilon_z(\omega)<0$. Since we have considered semi-infinite hBN crystal slabs, no Fabry-P\'erot hyperbolic phonon polariton modes~\cite{tomadin_prl_2015} appear in the energy loss function inside the reststrahlen bands.

\section{Computational details}
\label{app:technical}

The band structure calculations have been carried out by employing a plane-wave expansion of the Hamiltonian \eqref{eq:abstract_full_hamiltonian}. At each wave vector $\bm{k}$, we have used a basis of $271$ plane waves lying in the first $10$ hexagonal shells spanned by the moir\'e reciprocal lattice vectors. The total number of states in the basis was thus $271 \times 4 = 1084$, where the factor of $4$ comes from sublattice and layer indexes. We have computed the full spectrum but retained only half of it, i.e.~$\approx 500$ energy bands around the CNP.

The self-consistent solutions of Eqs.~(\ref{eq:self_consistency:hamiltonian})-(\ref{eq:self_consistency:density}) have been obtained with an absolute tolerance of $10^{-8}$ and a relative tolerance of $10^{-5}$. The self-consistency equations (\ref{eq:self_consistency:hamiltonian})-(\ref{eq:self_consistency:density}) have been solved explicitly with a Broyden iteration~\cite{johnson_prb_1988} (as in our calculations, a simple Anderson mixing iterative procedure did not converge to any solution).

For the optical conductivity and loss function, the integrals were performed over a mesh of $60\times60 = 3600$ equally spaced points in the MBZ. The value of $\eta$ was taken to be $\eta = 5~ {\rm meV}$.
\twocolumngrid

\end{document}